%% file: main.tex
\documentclass[fleqn,usenatbib]{mnras}

\usepackage{newtxtext,newtxmath}

\usepackage[T1]{fontenc}
\usepackage{ae,aecompl}

\usepackage{graphicx}	
\usepackage{amsmath}	
\usepackage{amssymb}	
\usepackage{threeparttable}
\usepackage{multirow}
\usepackage{url}
\usepackage{multicol}
\usepackage{xcolor}
\usepackage{subfig}

\usepackage{etoolbox}
\makeatletter
\patchcmd\@combinedblfloats{\box\@outputbox}{\unvbox\@outputbox}{}{%
   \errmessage{\noexpand\@combinedblfloats could not be patched}%
}%
\makeatother

\newcommand{\mum}{\ifmmode{\rm \mu m}\else{$\mu$m}\fi}

\newcommand{\chisq}{\ifmmode{\chi^{2} }\else{$\chi^2$}\fi}
\newcommand{\rchisq}{\ifmmode{\chi^{2} }\else{$\chi^2_\nu$}\fi}

\newcommand{\Hii}{H\,{\sc ii}}

\defcitealias{Robitaille2006}{R06} 
\defcitealias{Robitaille2017}{R17} 

\title[NGC 6822 YSOs]{The Young Stellar Population of the metal-poor galaxy NGC 6822}

\author[O. C. Jones et al.]{Olivia~C.~Jones,$^{1}$\thanks{E-mail: olivia.jones@stfc.ac.uk}
Michael J. Sharp,$^{2}$
Megan Reiter,$^{1}$
Alec S.\ Hirschauer,$^{3}$
M.~Meixner,$^{3}$
\newauthor
Sundar Srinivasan$^{4}$
\\
$^{1}$UK Astronomy Technology Centre, Royal Observatory, Blackford Hill, Edinburgh, EH9 3HJ, UK\\\(\)
$^{2}$Institute for Astronomy, University of Edinburgh, Blackford Hill, Edinburgh EH9 3HJ, UK\\\(\)
$^{3}$Space Telescope Science Institute, 3700 San Martin Drive, Baltimore, MD 21218, USA\\\(\)
$^{4}$Instituto de Radioastronomia y Astrofisica, UNAM Campus Morelia, Apartado Postal 3-72, 58090, Morelia, Michoacan, Mexico\\
}

\date{Accepted XXX. Received YYY; in original form ZZZ}

\pubyear{2019}

\begin{document}
\label{firstpage}
\pagerange{\pageref{firstpage}--\pageref{lastpage}}
\maketitle

\begin{abstract}
We present a comprehensive study of massive young stellar objects (YSOs) in the metal-poor galaxy NGC 6822 using IRAC and MIPS data obtained from the {\em Spitzer Space Telescope}. We find over 500 new YSO candidates in seven massive star-formation regions; these sources were selected using six colour-magnitude cuts. Via spectral energy distribution fitting to the data with YSO radiative transfer models we refine this list, identifying 105 high-confidence and 88 medium-confidence YSO candidates. For these sources we constrain their evolutionary state and estimate their physical properties. The majority of our YSO candidates are massive protostars with an accreting envelope in the initial stages of formation.  
We fit the mass distribution of the Stage I YSOs with a Kroupa initial mass function and determine a global star-formation rate of 0.039 $M_{\odot} yr^{-1}$.  This is higher than star-formation rate estimates based on integrated UV fluxes.
The new YSO candidates are preferentially located in clusters which correspond to seven active high-mass star-formation regions which are strongly correlated with the 8 and 24 $\mu$m emission from PAHs and warm dust. This analysis reveals an embedded high-mass star-formation region, Spitzer I, which hosts the highest number of massive YSO candidates in NGC 6822. The properties of Spitzer I suggest it is younger and more active than the other prominent H\,{\sc ii}  and star-formation regions in the galaxy. \end{abstract}




\section{Introduction}
\label{sec:intro}

NGC 6822 is an isolated gas-rich barred irregular galaxy in the Local Group. At a distance of $d = 490\pm 40$ kpc \citep{Sibbons2012,Sibbons2015}, it is the third-nearest dwarf irregular galaxy after the Large and Small Magellanic Clouds (LMC and SMC respectively); and has a metallicity comparable to that of the SMC ($\sim0.2~{\rm Z}_{\odot}$; \citealt{Skillman1989, Lee2006a, GarciaRojas2016}). Its proximity, low metallicity, and lack of known close companions \citep{deBlok2000} makes NGC 6822 an ideal candidate for studying resolved stellar populations in an undisturbed system. 
We summarise the global properties of NGC 6822 in Table~\ref{tab:ngc6822Basics}. 

NGC 6822 has many prominent metal-poor H\,{\sc ii} regions and OB associations \citep{Efremova2011, Rubin2016, Schruba2017} which are actively forming massive stars. 
Despite these being amongst the brightest and most massive star-forming regions known, this isolated galaxy has a low global star-formation rate (SFR) of $\sim0.06 ~{\rm M}_{\odot} \, {\rm yr}^{-1}$ \citep{Israel1996, Efremova2011}, which appears to have been relatively consistent over the last 11 Gyr. Recently, the SFR has increased \citep{Wyder2003, Cannon2012}, particularly in the past 400 Myr \citep{Gallart1996b, Clementini2003, Komiyama2003}, with a burst of star formation in the bar $<$200 Myr ago \citep{deBlok2000}. 
The young stellar content of NGC 6822 appears mainly in a central bar structure \citep{Schruba2017}, which is orientated in a north-south direction in the inner part of the  H\,{\sc i} distribution, whilst the old- and intermediate-age population is elliptically distributed \citep{Demers2006, Letarte2002}, and dynamically decoupled from the H\,{\sc i} envelope \citep{Battinelli2006}.

Despite detailed studies of the prominent star-forming regions in NGC 6822, its massive young stellar population has been more difficult to characterise on a global scale, and no global survey of its intermediate- to high-mass young stellar objects (YSOs) and how they relate to the galaxy's gas and dust distributions have been conducted. 
In this paper we investigate the massive young stellar population of NGC 6822 to infer the properties of metal-poor star formation without the additional influence of tidal effects from any associated neighbours (as in the SMC). In Section~\ref{sec:data} we describe the photometric data and how we selected the YSO candidates. 
Given the distance of NGC 6822 and the resolution of {\em Spitzer}, these candidates are unlikely to be individual sources, but proto-clusters which are dominated by the most luminous source \citep{Chen2009, Oliveira2009, Ward2017}. This technique has been effective in the LMC and SMC to identify embedded regions that are actively forming stars over the last 0.2 Myr, as apposed to other tracers e.g., H$\alpha$ and UV emission from massive stars, which identify star-formation on $\sim$10 Myr timescales \citep{Whitney2008, Sewilo2013}.
In Section~\ref{sec:Models} we perform spectral energy distribution (SED) fitting with YSO models and discuss the results in Sections \ref{sec:results} and~\ref{sec:discussion}. Finally, our conclusions are summarised in Section~\ref{sec:conclusion}.

\begin{table}
\centering
\begin{minipage}{84mm}
 \caption{Summary of the Global Properties of NGC 6822}
 \label{tab:ngc6822Basics}
\centering
 \begin{tabular}{@{}lcl@{}}
   \hline
   \hline
Property   &   Value    &   References  \\          
\hline             
 Right ascension    &  19 44 56.4        &        \\        
 Declination        & -14 48 04.5        &         \\                
 Distance           &  $490\pm 40$       &    \citet{Rich2014}        \\               
  $(m - M)_0$       &   $23.34\pm 0.06$  &    \citet{Pietrzynski2004}                \\   
 Position angle     &  $115\pm 15$ deg   &    \citet{Weldrake2003}     \\          
 Inclination        &  $60\pm 15$  deg   &    \citet{Weldrake2003}     \\
 Systemic vel.      &   $-57 \pm 2$ km s$^{−1}$       &    \citet{Koribalski2004}     \\    
 E(B -- V)          &  0.21              &    \citet{Schlafly2011}     \\  
 12 + log(O/H)      & $8.02\pm 0.05$  &    \citet{GarciaRojas2016}      \\   
 M$_{\rm V}$        & -15.96          &    \citet{Dale2007}     \\                
 ${M}_{\mathrm{star}}$    &  $1.5\times {10}^{8}$ ${M}_{\odot }$              &  \citet{Madden2013}     \\   
 ${M}_{\mathrm{dust}}$    &  $3\times {10}^{5}$ ${M}_{\odot }$                &  \citet{RemyRuyer2015}   \\   
 Gas-to-dust ratio        &    $\sim$186 -- 480         &  \citet{Zubko2004}     \\ 
                          &                             &   \citet{Galametz2010}      \\           
       
 SFR UV                  &   0.015 ${\rm M}_{\odot} \, {\rm yr}^{-1}$   &    \citet{Efremova2011}       \\           
  \hline
 \end{tabular}
\end{minipage}
\end{table}

\section{Photometric Data}
\label{sec:data}

\subsection{Point Source Catalogues}

There is excellent wide-field coverage of NGC 6822 in the optical to mid-IR. 
We use these deep, uniform photometric catalogues to produce a combined multi-wavelength data set from which we identify YSO candidates and assess potential contamination from other populations of sources. 

Near-IR $JHK$ photometry covering an area of 3 deg$^2$ was taken from  \cite{Sibbons2012}. They observed NGC 6822  with the Wide Field CAMera (WFCAM) on the 3.8 m United Kingdom Infrared Telescope (UKIRT). This data is complete to a depth of 19.5 mag in the $J$-band and 18.7 in $K$.   
In the mid-IR, NGC 6822 was observed with {\em Spitzer} Infrared Array Camera (IRAC) and Multiband Imaging Photometer for {\em Spitzer} (MIPS) as part of the {\em Spitzer} Infrared Nearby Galaxies Survey \citep[SINGS;][]{Kennicutt2003}. This data covers the central $\sim$0.1 deg$^2$ area of NGC 6822 and encompasses the optical emission from the galaxy to the R$_{25}$ level \citep{Cannon2006}. 
This data was processed by \citet{Khan2015} using aperture photometry to generate a {\em Spitzer} point-source catalogue of 30,745 IRAC objects of which 7,268 had a MIPS 24 $\mu$m counterpart. 

The {\em Spitzer} data are matched to the near-IR ($JHK$) data via positional cross-matching using a radius of 1$\arcsec$; the closest match was selected if multiple stars met this criteria. To ensure a high photometric reliability in the cross-matched catalogue only stars with standard errors less than  0.1 mag were included in the matching process. See Hirschauer et al in prep; Paper I for more details on the generation of the master catalogue. The initial cross-matched catalogue contains 30,745 sources in a 0.115  deg$^2$ region. 
To construct a reliable list of YSO candidates we require that each source has valid fluxes in at least three of the five {\em Spitzer} bands or at least two of the three {\em WFCAM} bands and two {\em Spitzer} IRAC detections. 
We also remove objects from the catalogue without a 3.6 $\mu$m or a 4.5 $\mu$m flux and those that have magnitude uncertainties  $>$0.2 mag, which leaves 23,908 sources, 78\% of the initial catalogue. 

\subsection{Colour classifications}

In order to identify candidate YSOs in NGC 6822 we adapt the successful methodology applied on a galaxy-wide scale by \citet{Whitney2008} and \citet{Sewilo2013} to {\em Spitzer} IRAC and MIPS data for the Magellanic Clouds \citep{Meixner2006, Gordon2011}. This strict multi-dimensional colour-magnitude selection criteria (using five or more different CMDs) developed from theoretical models and combined with spectral energy distribution (SED) fitting with YSO models was shown to be  $\sim$80\% reliable for complex populations and over 95\% accurate in verified star-forming regions \citep[e.g.][]{Carlson2012} by \cite{Jones2017b}. These selections separate YSO candidates from evolved stars and divide YSOs into two categories; a reliable candidate list and a possible YSO candidate list, based on their near- and mid-IR colours.

Our combined point-source catalogue contains a total of 30,745 stars, 14,486 of which have near-IR counterparts and 16,259 of which only appear in the mid-IR.  Due to dust extinction, heavily embedded protostars may be too faint in the near-IR to be detected, and only mid-IR photometry can recover these sources. Out of the 20,999 reliable {\em Spitzer} sources in NGC 6822, 4,153 have a MIPS 24~$\mu$m counterpart. These very red sources are indicative of young YSOs and are traditionally classified as Class I or Stage I sources (based on their IR spectral index or mass-accretion rates, respectively; \citealt{Lada1987, Robitaille2006}). 

\begin{figure*}
    \centering
    \vspace{0 cm}
    \subfloat[\label{subfig:36_58_Hess}]{\includegraphics[width = 150 pt]{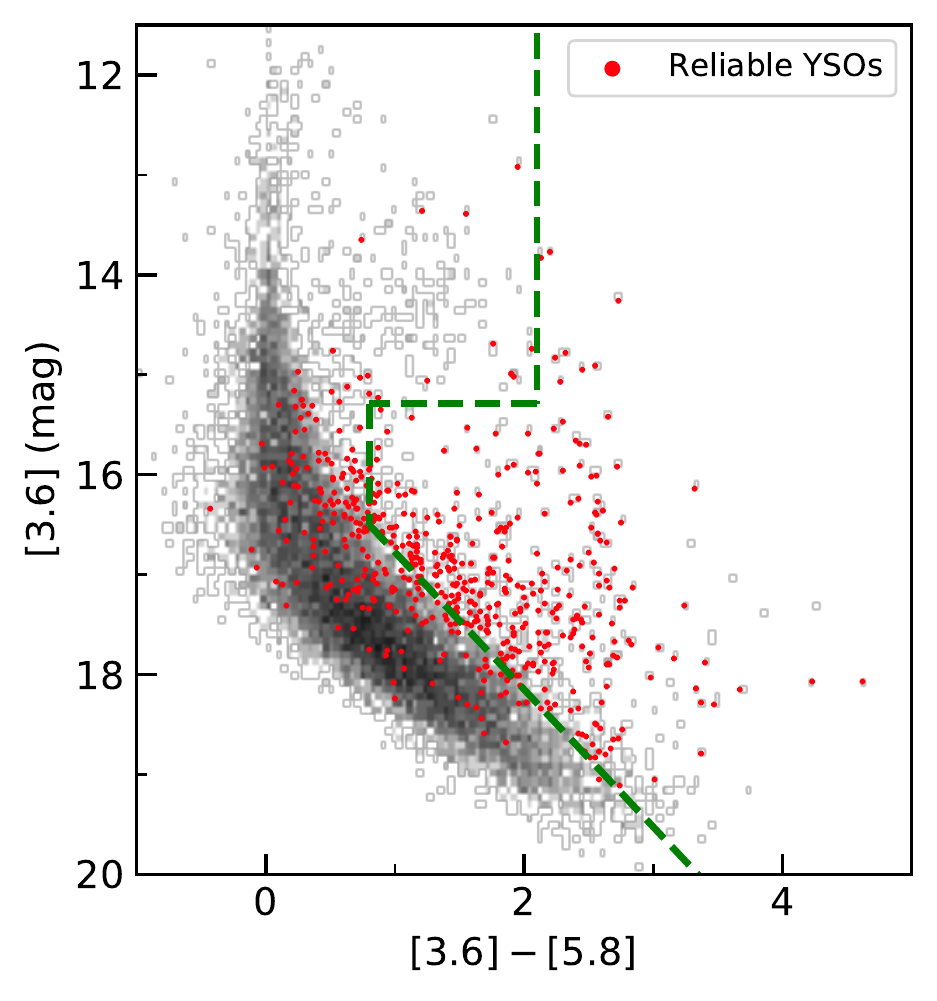}}
    \subfloat[\label{subfig:45_58_Hess}]{\includegraphics[width = 150 pt]{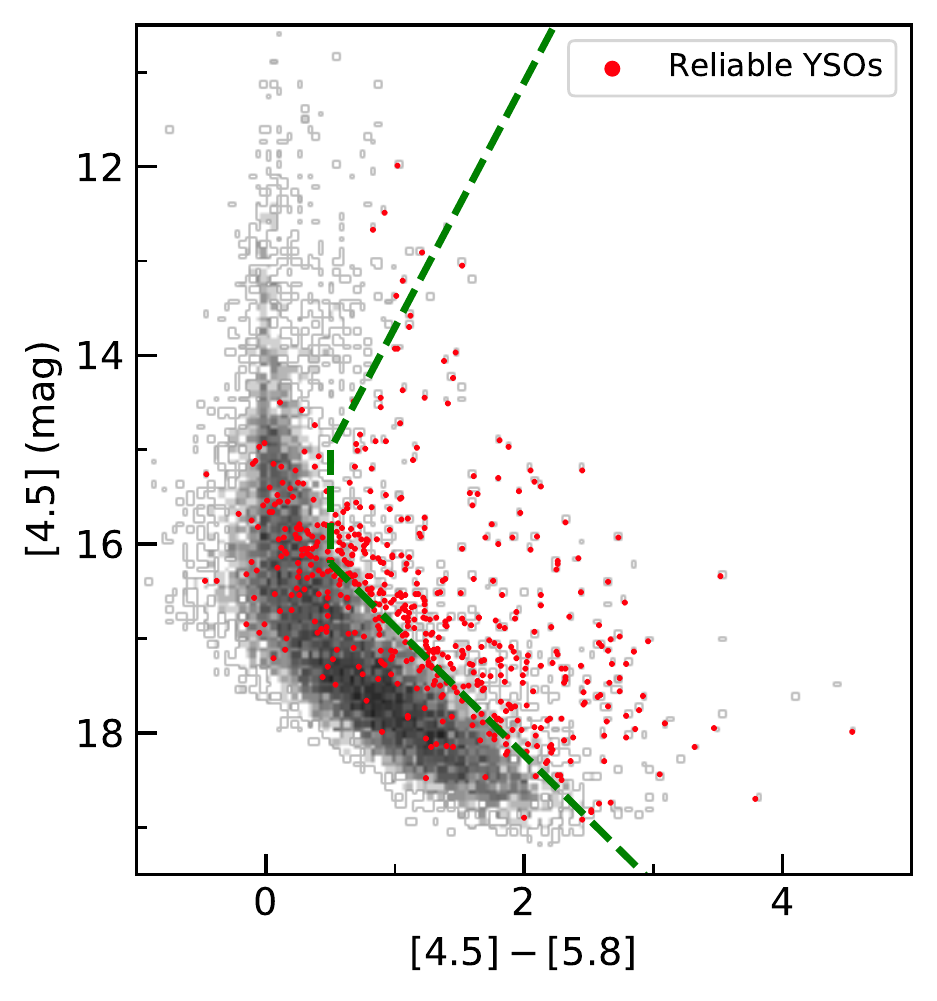}}
    \subfloat[\label{subfig:45_24_Hess}]{\includegraphics[width = 150 pt]{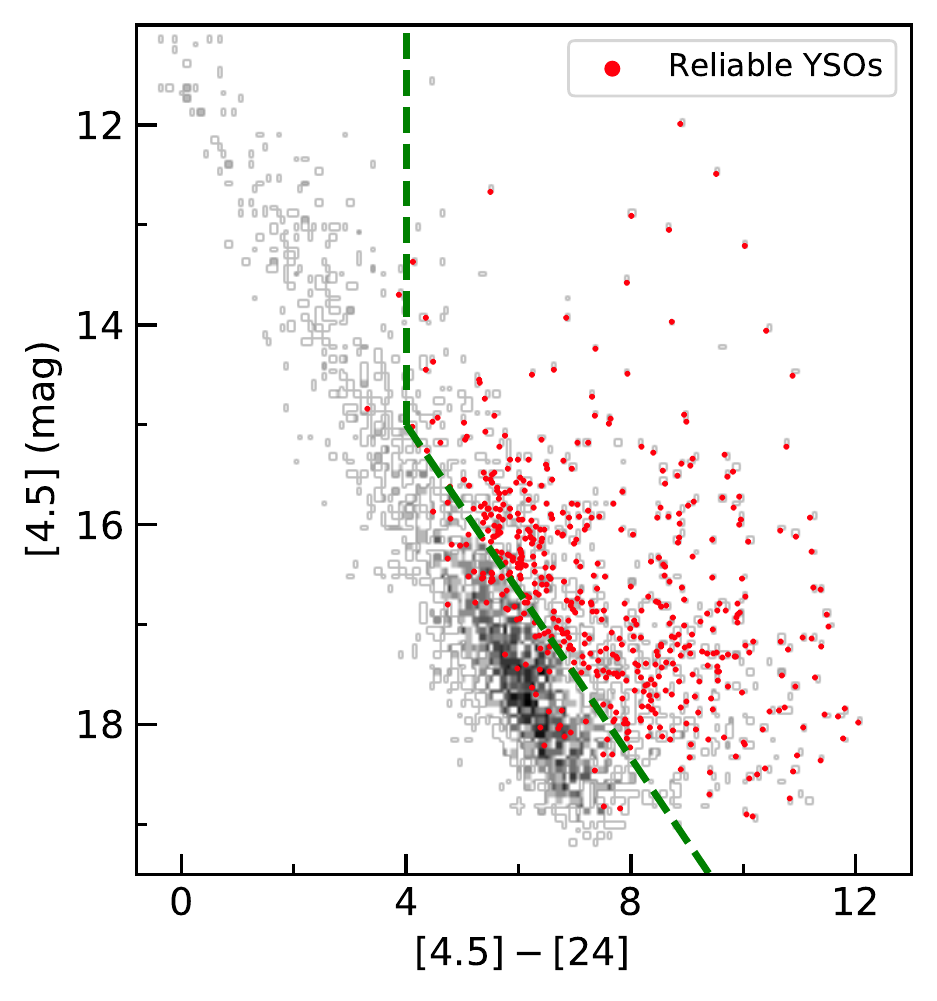}} \\
    \subfloat[\label{subfig:36_80_Hess}]{\includegraphics[width = 150 pt]{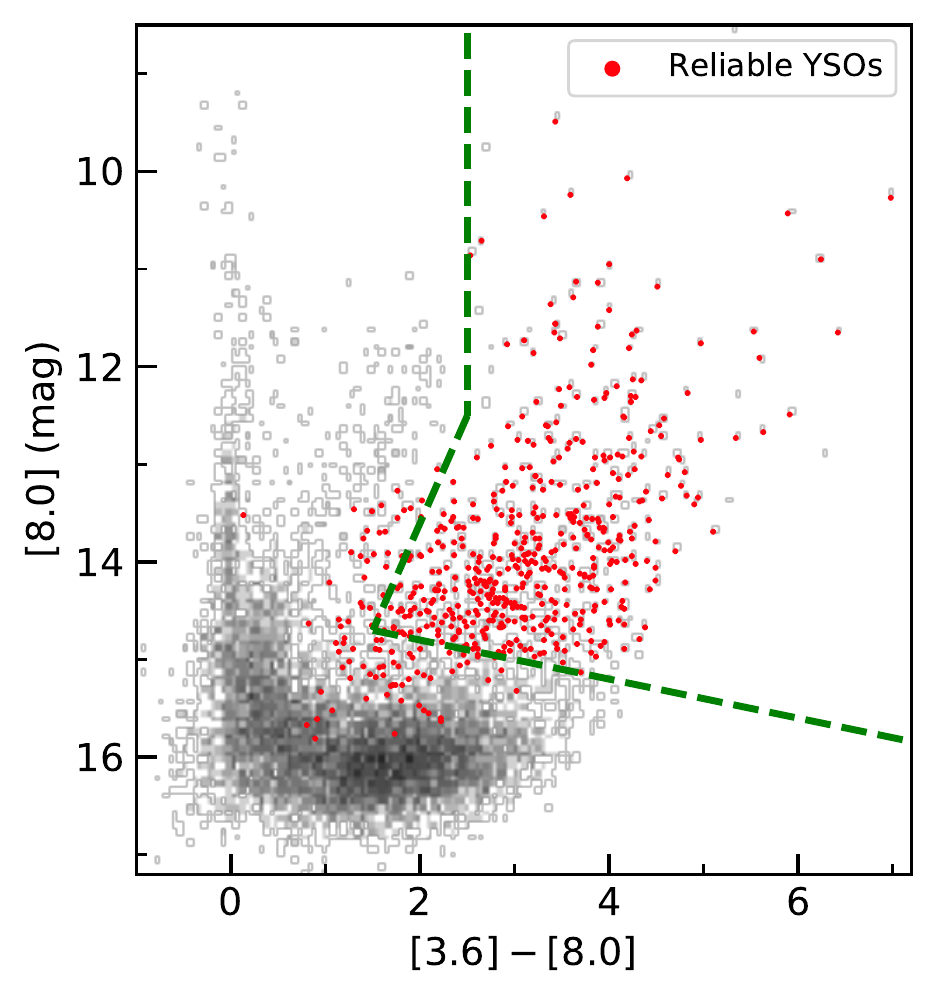}}
    \subfloat[\label{subfig:45_80_Hess}]{\includegraphics[width = 150 pt]{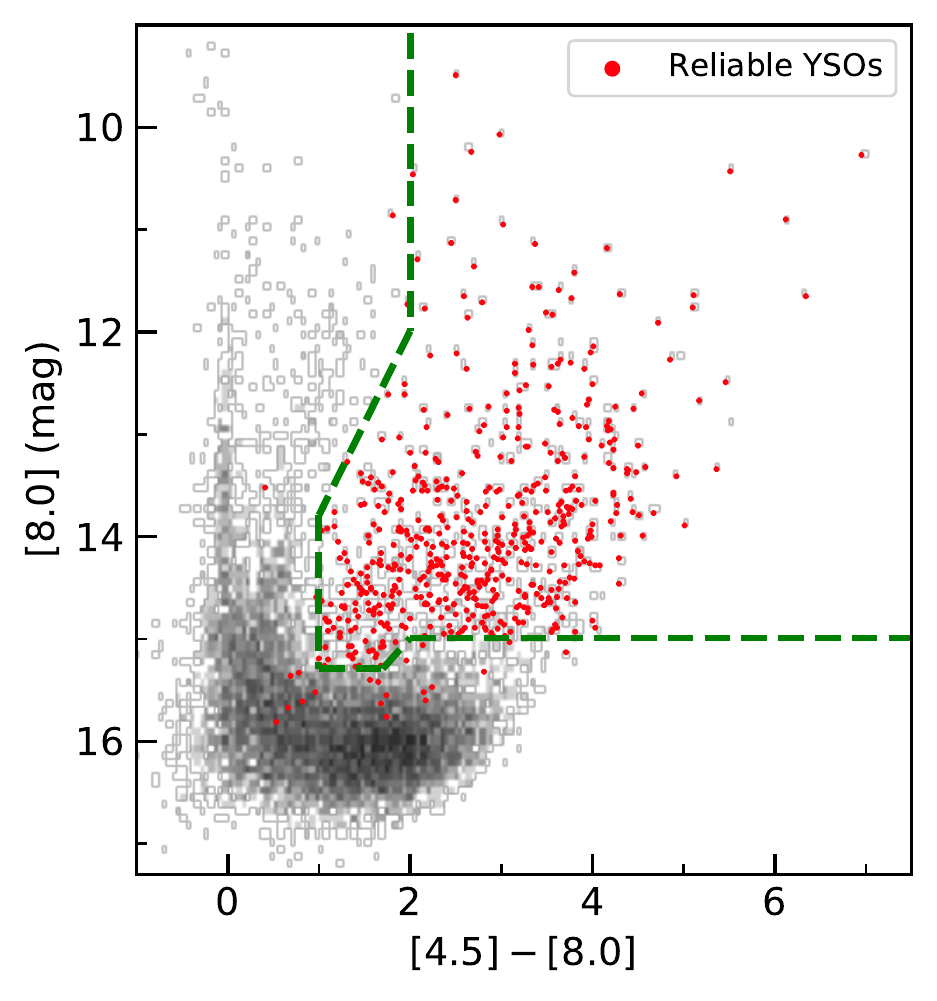}}
    \subfloat[\label{subfig:80_24_Hess}]{\includegraphics[width = 150 pt]{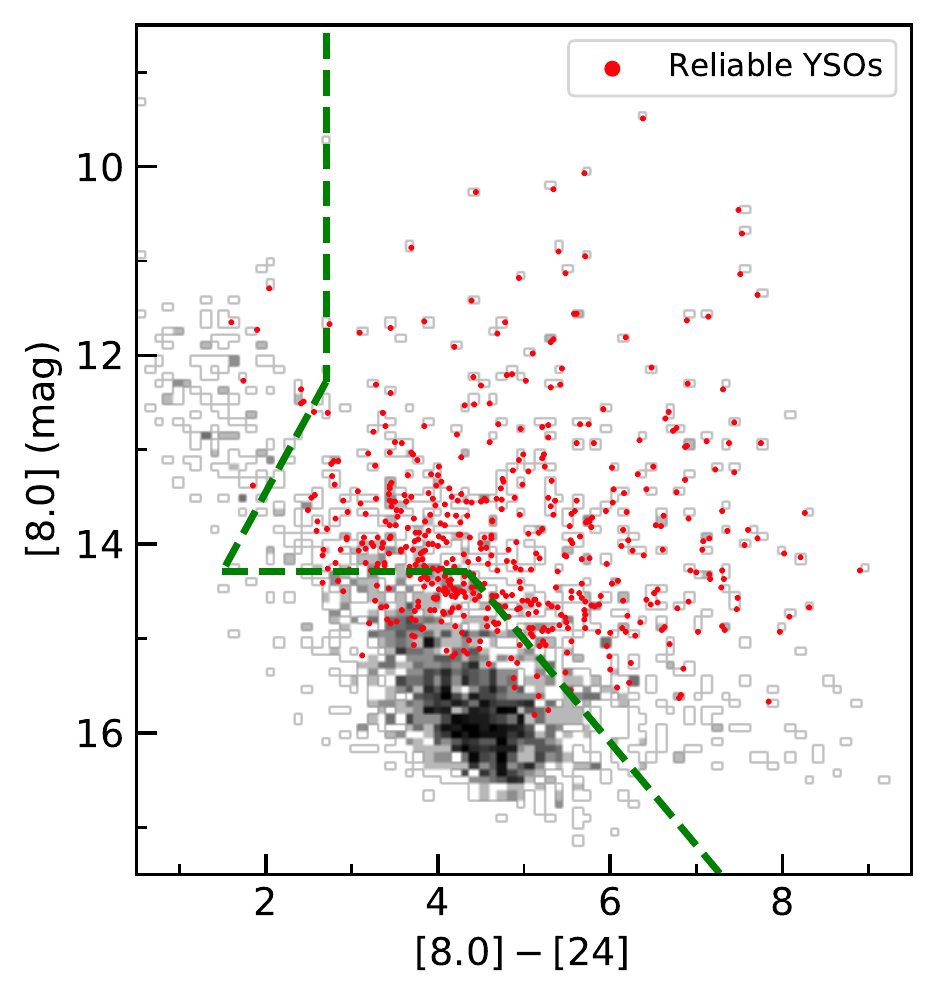}}
    \caption{Mid-IR CMD for point sources in NGC 6822, showing the separation of YSOs from other stellar populations. The CMDs are represented as Hess diagrams in grey scale, our reliable YSO candidates are shown as red points.  The green dashed lines denote the various cuts used to generate YSO candidate lists given in Equations~\ref{eq:I1I3_cuts}--\ref{eq:I4M1_cuts}, they represent the boundaries between regions occupied by candidate YSOs and other non-YSO populations. The CMDs shown are: [3.6] vs.~[3.6]--[5.8], [4.5] vs.~[4.5]--[5.8], [4.5] vs.~[4.5]--[24], [8.0] vs.~[3.6]--[8.0], [8.0] vs.~[4.5]--[8.0] and [8.0] vs.~[8.0]--[24].}
    \label{fig:CMDs}
\end{figure*}

Figure~\ref{fig:CMDs} shows the stellar-density distribution of NGC 6822 in the six mid-IR CMDs used to select candidate YSOs. The CMDs are displayed as Hess diagrams, with brightness of each pixel corresponding to the number density of sources. Different populations can be identified in the CMDs. To select the YSO candidates, we first identify sources with an IR-excess. 
This red population (37\% of the original catalogue) is comprised of YSOs, evolved stars, reddened stellar photospheres, and background galaxies, all of which can overlap significantly in colour-magnitude space. A series of colour-magnitude selection criteria, adjusted for the distance to NGC 6822, given in Equations~\ref{eq:I1I3_cuts}--\ref{eq:I4M1_cuts} and shown graphically in Figure~\ref{fig:CMDs} are then used to isolate sources in regions of colour-magnitude space occupied predominantly by YSOs and remove contaminating sources.  We do not correct the colours for reddening due to dust along the line of sight since, at mid-IR wavelengths, the reddening is small.
To account for the difference in distance between the SMC and NGC 6822 we adjust the boundaries of the magnitude cuts from \citet{Sewilo2013} to be  4.39 mag fainter. For NGC 6822 we adopt a distance modulus of $\mu = 23.34$ mag \citep{Pietrzynski2004}. Similarly, we adopt $\mu = 18.96$ mag for the SMC \citep{deGrijs2015}. 


Altering the faint-source limit in the mid-IR selection criteria may result in the selection of sources in the galaxy-dominated region of the CMD. We test this by comparing the distance-adjusted cuts from \citet{Whitney2008} and \citet{Sewilo2013} to the source densities in the respective Hess diagrams. For the $[3.6] - [5.8]$ and $[4.5] - [5.8]$ CMDs the adjusted colour selection becomes unreliable at faint magnitudes due to the large overlap YSOs in NGC 6822 have with the general and background populations.  To ensure a reliable selection of candidate YSOs, we increase the faint-source limits for these CMDs to minimise contamination, at the cost of reduced completeness.


\small
\renewcommand{\arraystretch}{0.85}
\begin{eqnarray}
\label{eq:I1I3_cuts} 
\left \lbrace
\begin{array}{r}
        [3.6] < 15.29  \; \; {\rm and} \; \; [3.6] - [5.8] \geqslant 2.1 \; \;{\rm or} \\
 \\
   15.29 < [3.6] < 16.5  \; \; {\rm and} \; \; [3.6] - [5.8] \geqslant 0.8 \; \;{\rm or} \\
 \\ 
 ~16.5 \leqslant [3.6]  < (15.4+1.375 \cdot ([3.6] - [5.8]))  \; \; {\rm and} \\\; \; [3.6] - [5.8] > 0.8\; \; \; \;\; \\
\end{array}     \right \rbrace 
\end{eqnarray}
\normalsize
\renewcommand{\arraystretch}{1}

\small
\renewcommand{\arraystretch}{0.85}
\begin{eqnarray}
\label{eq:I2I3_cuts} 
\left \lbrace
\begin{array}{r}
        [4.5]  \leqslant 15.0  \; \; {\rm and} \; \; [4.5] > (16.53-2.59 \cdot ([4.5] - [5.8])) \; \;{\rm or} \\
 \\        
        ~15.0 < [4.5]  < 16.2    \; \; {\rm and}  \; \; [4.5] - [5.8] \geqslant 0.5 \; \;{\rm or} \\
  \\       
   ~16.2 \leqslant [4.5] < (15.5 + 1.36   \cdot  ([4.5] - [5.8])) \;\; \;\;\; \\
\end{array}     \right \rbrace 
\end{eqnarray}
\normalsize
\renewcommand{\arraystretch}{1}


\small
\renewcommand{\arraystretch}{0.85}
\begin{eqnarray}
\label{eq:I2M1_cuts} 
\left \lbrace
\begin{array}{r}
        [4.5]  < 15.0  \; \; {\rm and} \; \; [4.5] - [24] \geqslant 4.0 \; \;{\rm or} \\
 \\        
        ~[4.5]  \geqslant 15.0  \; \; \; \; {\rm and} \; \; \; \; [4.5] < (11.67+0.833 \cdot ([4.5] - [24])) \; \; \; \;  \;\\ 

\end{array}     \right \rbrace 
\end{eqnarray}
\normalsize
\renewcommand{\arraystretch}{1}


\small
\renewcommand{\arraystretch}{0.95}
\begin{eqnarray}
\label{eq:I1I4_cuts} 
\left \lbrace
\begin{array}{r}
        [8.0]  < 12.5  \; \; {\rm and} \; \; [3.6] - [8.0] \geqslant 2.5 \; \;{\rm or} \\
 \\        
        ~12.5 \leqslant [8.0]  \leqslant 14.7    \; \; {\rm and}  \; \; [8.0] > (18.0-2.2 \cdot ([3.6] - [8.0])) \; \;{\rm or} \\
  \\       
   ~14.7 < [4.5] \leqslant 16.0  \; \; {\rm and}  \; \; [8.0] < (14.4+0.2 \cdot ([3.6] - [8.0]))    \;\; \;\;\; \\

\end{array}     \right \rbrace 
\end{eqnarray}
\normalsize
\renewcommand{\arraystretch}{1}


\small
\renewcommand{\arraystretch}{0.95}
\begin{eqnarray}
\label{eq:I2I4_cuts} 
\left \lbrace
\begin{array}{r}
        [8.0]  < 14.99  \; \; {\rm and} \; \; [4.5] - [8.0] > 2.0 \; \;{\rm or} \\
 \\        
(15.59-1.8 \cdot ([4.5] - [8.0]))< [8.0] \leqslant 15.29   \; \; {\rm and}  \\
~1 \leqslant [4.5] - [8.0] < 1.7 \; \;{\rm or} \\
\\
~[8.0] \leqslant 16.59-([4.5] - [8.0]))   \; \; {\rm and}  \\
~[8.0] \geqslant 15.59-1.8 \cdot ([4.5] - [8.0]))   \; \; {\rm and}  \\
~1.7 \leqslant [4.5] - [8.0] \leqslant 2.0\; \;  \; \; \; \; \; \\
\end{array}     \right \rbrace 
\end{eqnarray}
\normalsize
\renewcommand{\arraystretch}{1}


\small
\renewcommand{\arraystretch}{0.95}
\begin{eqnarray}
\label{eq:I4M1_cuts} 
\left \lbrace
\begin{array}{r}
~[8.0]  < 14.29   \; \; {\rm and} \; \; [8.0]  - [24] \geqslant 2.7 \; \;{\rm or} \\
 \\
 ~[8.0]  < 14.29   \; \; {\rm and} \; \; [8.0] - [24] \leqslant 2.7 \; \;  \; \; {\rm and}\\
  ~[8.0] > (16.79-1.67 \cdot ([8.0] - [24]))\; \;\; {\rm or} \\
 \\   
  ~[8.0] \geqslant 14.29 \;\; {\rm and} \; \; [8.0] < (9.56 + 1.09  \cdot  ([8.0]  - [24])) \; \;\; \; \\
\end{array}     \right \rbrace 
\end{eqnarray}
\normalsize
\renewcommand{\arraystretch}{1}

Mid-IR colours with longer photometric baselines  (e.g.~[3.8]-[8.0], [4.5]-[8.0] and [8.0]--[24]) are best at separating YSOs into a distinct group. In these bands YSOs appear extremely red as their SEDs rise steeply toward longer wavelengths. 
The 24 $\mu$m data point helps constrain the long-wavelength shape of the SED and therefore the luminosity of the sources.  Short baselines are less effective; colours of YSOs may mirror those exhibited by dust features from other classes resulting in a degeneracy or, in some instances, be masked by the noise for faint sources.

\begin{figure}
    \centering
    \vspace{0 cm}
    \includegraphics[width=0.45\textwidth]{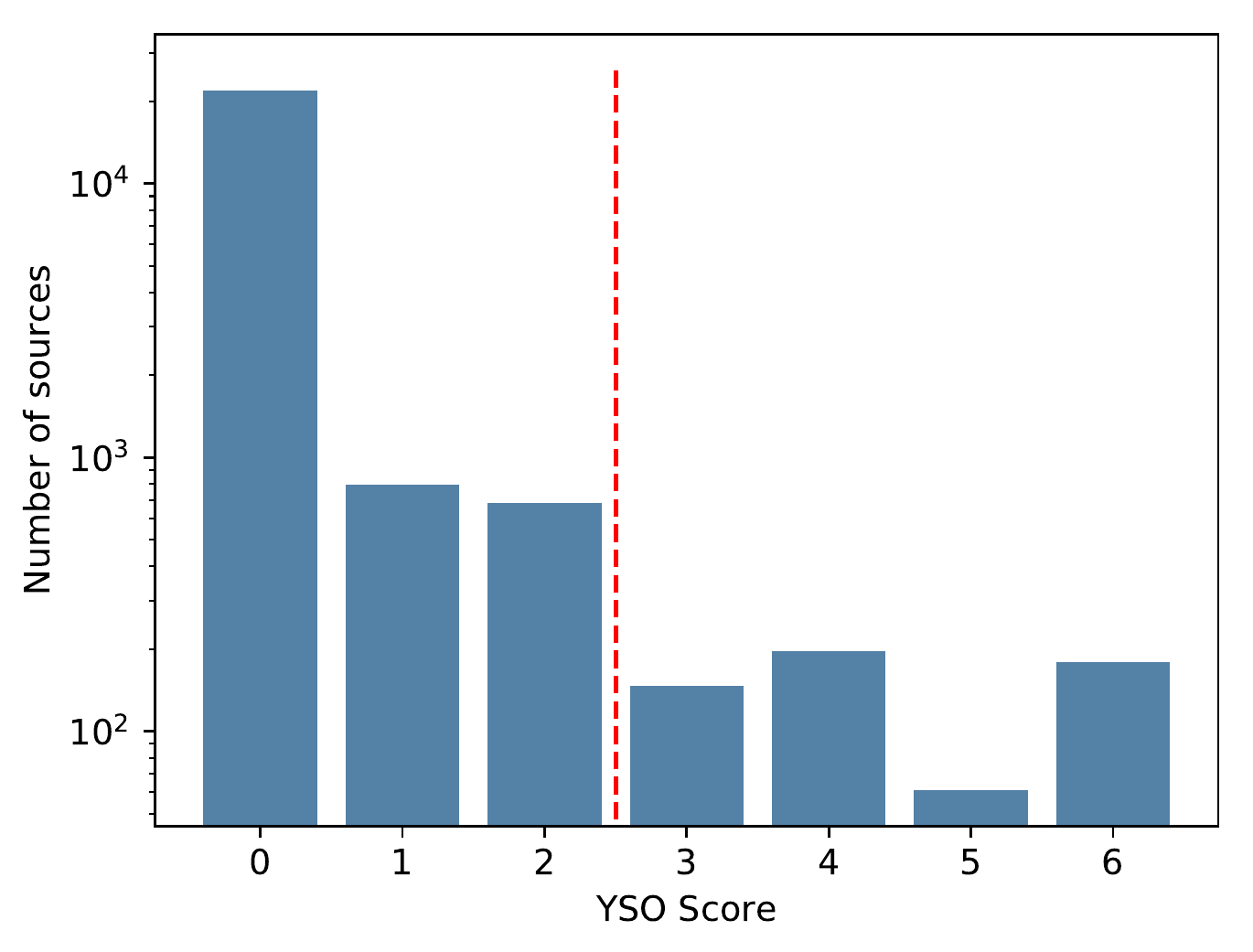}
    \caption{CMD score distribution for the NGC 6822 population. The dashed line indicates the divide between non-YSOs and YSOs candidates in our initial selection.  Only objects with a score of three or more were considered for further analysis. }
    \label{fig:cmdscore_dist}
\end{figure}

We assign each source a colour score based on the total number of CMDs in which it meets the YSO colour-criteria, the distribution of scores in shown in  Figure~\ref{fig:cmdscore_dist}.  
Sources with a score of three or more are `reliable' YSO candidates; this selection contains 584 sources. We consider the 680 sources with a score of two to be `possible' YSO candidates. 
Tables~\ref{tab:ngc6822_HP_YSOtable} and~\ref{tab:ngc6822_YSOtable_prob} contains the list of positions, $JHK$, IRAC, and MIPS [24] magnitudes, and uncertainties for the  584 reliable YSO candidates in NGC 6822. These have been separated into a high-confidence and a medium-confidence YSO list based on their SED fit in Section~\ref{sec:sedfit}.

Our data are only sensitive to massive ($>$8${M}_{\odot }$) sources, so the YSO sample included in this work is not complete, due to decreasing completeness of the {\em Spitzer} data with increasing wavelength. Colour-magnitude selections are biased towards brighter sources, as these have both a cleaner separation for low-luminosity background galaxies in the CMDs and better S/N ratios.  
Multiple colour criteria can be utilised to assign a classification, making the identification more reliable. 
Moreover, relying on mid-IR selection criteria biases our YSO candidate list towards young, embedded sources of intermediate and high mass at the initial stages (Stage I and II ) of formation \citep{Whitney2008}. More evolved (Stage III) YSOs, which have hot central objects, have a greater degeneracy in colour-magnitude space (e.g., between YSOs, evolved stars, and galaxies), which cannot be resolved with broad band infrared photometry.

\begin{table*}
\centering
 \caption{Table of photometry for the high-confidence YSO Candidates. The CMD score, cluster membership, and if there is PAH enhancement, or possibly a stellar photosphere present is also recorded.  Only a portion of this table is shown here to demonstrate its form and content. A machine-readable version of the full table is available online.}
 \label{tab:ngc6822_HP_YSOtable}
\centering
\scalebox{0.8}{
 \begin{tabular}{@{}ccccccccccccccc@{}}
   \hline
   \hline
      YSO  &    R.A.  &    Dec.  &  	WFCAM  & 	WFCAM  & 	WFCAM  & 	IRAC  & 	IRAC  & 	IRAC  & 	IRAC  & 	MIPS  &  CMD  & PAH            & Stellar         & Cluster \\	
       ID  &          &          &	$J$    &     $H$  & 	$K_s$  & 	[3.6]  & 	[4.5]  & 	[5.8]  & 	[8.0]  & 	[24]      &	score          & enhancement   &  photosphere  &    \\	
\hline             
    162 & 296.1993 & -14.8757 & -99.99 & -99.99 & -99.99 & 17.92 & 17.83 & 16.22 & 14.57 & 7.09 & 4 & 1 & 0 & 2 \\
    179 & 296.2022 & -14.7391 & -99.99 & -99.99 & -99.99 & 17.09 & 17.13 & 15.61 & 14.50 & 8.95 & 6 & 1 & 0 & -1 \\
    188 & 296.2046 & -14.7163 & -99.99 & -99.99 & -99.99 & 15.91 & 15.44 & 13.48 & 11.67 & 8.93 & 6 & 1 & 0 & 3 \\
    189 & 296.2046 & -14.8818 & -99.99 & -99.99 & -99.99 & 17.89 & 18.05 & 15.67 & 13.85 & 7.70 & 6 & 1 & 0 & 2 \\
    191 & 296.2054 & -14.8830  & -99.99 & -99.99 & -99.99 & 16.99 & 16.54 & 14.41 & 12.90 & 6.56 & 6 & 1 & 0 & 2 \\
  \hline
 \end{tabular}}
\end{table*}

\begin{table*}
\centering
 \caption{Table of photometry for the medium-confidence YSO Candidates. The CMD score, cluster membership, and if there is PAH enhancement, or possibly a stellar photosphere present is also listed. Only a portion of this table is shown here to demonstrate its form and content. A machine-readable version of the full table is available online.}
 \label{tab:ngc6822_YSOtable_prob}    
\centering
\scalebox{0.8}{
 \begin{tabular}{@{}ccccccccccccccc@{}}
   \hline
   \hline
   YSO  &    R.A.  &    Dec.  &  	WFCAM  & 	WFCAM  & 	WFCAM  & 	IRAC  & 	IRAC  & 	IRAC  & 	IRAC  & 	MIPS  &  CMD  & PAH            & Stellar    & Cluster \\	
     ID   &           &           &	$J$    &     $H$  & 	$K_s$  & 	[3.6]  & 	[4.5]  & 	[5.8]  & 	[8.0]  & 	[24]  &	 score & enhancement   &    photosphere &   \\		
\hline             
1 & 296.0293 & -14.6865 & 19.31 & 18.87 & 18.05 & 17.16 & 16.14 & 15.31 & 14.02 & 10.02 & 6 & 0 & 0 & -1 \\
7 & 296.0566 & -14.8798 & -99.99 & -99.99 & -99.99 & 17.01 & 16.05 & 15.69 & 15.36 & 9.88 & 3 & 0 & 1 & -1 \\
9 & 296.0590 & -14.7746 & -99.99 & -99.99 & -99.99 & 18.77 & 18.82 & 16.30 & 14.82 & 11.30 & 4 & 1 & 0 & -1 \\
11 & 296.0663 & -14.9373 & -99.99 & -99.99 & -99.99 & 16.82 & 15.80 & 15.40 & 14.55 & 10.07 & 4 & 0 & 1 & -1 \\
15 & 296.0681 & -14.7610 & -99.99 & -99.99 & -99.99 & 17.36 & 16.78 & 15.90 & 14.87 & 10.67 & 3 & 0 & 0 & -1\\
  \hline
 \end{tabular}}
\end{table*}

\begin{table*}
\centering
 \caption{Magnitudes and uncertainties for the possible YSO Candidates. Only a portion of this table is shown here to demonstrate its form and content. A machine-readable version of the full table is available online.}
 \label{tab:ngc6822_YSOtable_poss}    
\centering
\scalebox{0.7}{
 \begin{tabular}{@{}cccccccccccccc@{}}
   \hline
   \hline
 R.A.  &    Dec.  &  	WFCAM  & 	WFCAM  & 	WFCAM  & 	IRAC  & 	IRAC  & 	IRAC  & 	IRAC   & 	MIPS  &  CMD  & PAH     &    Stellar     \\	
       &          &	$J$    &     $H$  & 	$K_s$  & 	[3.6]  & 	[4.5]  & 	[5.8]  & 	[8.0]  & 	[24]  &	 score & enhancement &    photosphere    \\	
 \hline             
296.1919 & -14.9305 & 18.16 $\pm$ 0.06 & 17.53  $\pm$ 0.05 & 17.39 $\pm$  0.06 & 17.12 $\pm$  0.09 & 17.21 $\pm$  0.12 & 15.86 $\pm$  0.07 & 16.15 $\pm$  0.15 & 12.70 $\pm$  -99.99 & 2 & 0 & 0 \\
296.1927 & -14.9303 & 16.94 $\pm$ 0.02 & 16.28  $\pm$ 0.02 & 16.22 $\pm$  0.02 & 15.98 $\pm$  0.04 & 15.97 $\pm$  0.04 & 15.05 $\pm$  0.07 & 15.81 $\pm$  0.14 & 12.44 $\pm$   -99.99 & 2 & 0 & 0  \\
296.1930 & -14.9311 & -99.99 $\pm$ -99.99 & -99.99 $\pm$  -99.99 & -99.99$\pm$   -99.99 & 18.15 $\pm$  0.13 & 17.68 $\pm$  0.13 & 15.82 $\pm$  0.14 & 16.16  $\pm$  0.13 & 12.10 $\pm$  0.25 & 2 & 1 & 0 \\ 

296.1945 & -14.9341 & -99.99 $\pm$ -99.99 & -99.99 $\pm$  -99.99 & -99.99$\pm$   -99.99 & 18.04 $\pm$  0.12 & 18.28 $\pm$  0.16 & 16.02 $\pm$  0.11 & 15.9 $\pm$  -99.99 & 12.53 $\pm$  0.29 & 2 & 1 & -99.99 \\
296.1948 & -14.8460  & 19.73 $\pm$ 0.23 & 18.81 $\pm$ 0.16 & 17.53 $\pm$ 0.07 & 17.17  $\pm$ 0.07 & 16.66  $\pm$ 0.04 & 16.19  $\pm$ 0.08 & 13.99  $\pm$ 0.07 & 10.87 $\pm$ 0.07 & 3 & 0 & 1 \\
  \hline
 \end{tabular}}
\end{table*}



\begin{figure*}
    \centering
    \vspace{0 cm}
    \subfloat[\label{subfig:H_K_K_45_Hess}]{\includegraphics[width = 150 pt]{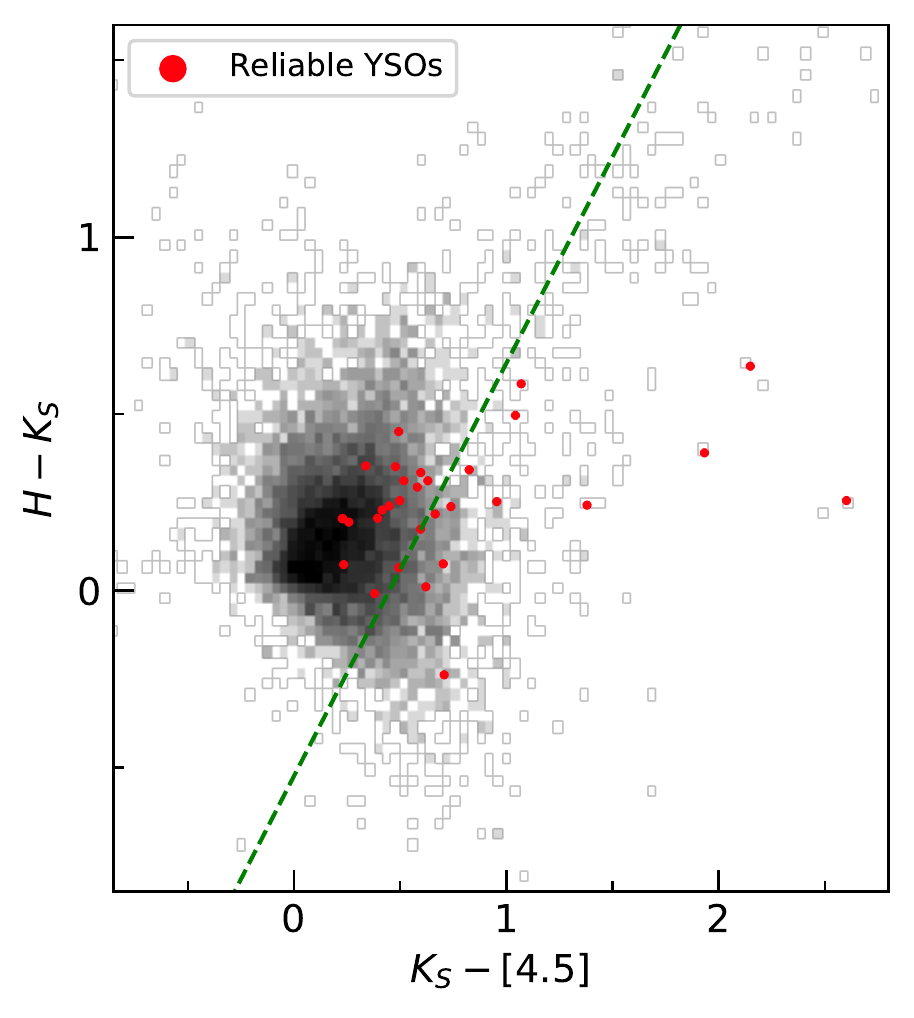}}
    \subfloat[\label{subfig:J_K_J_45_Hess}]{\includegraphics[width = 150 pt]{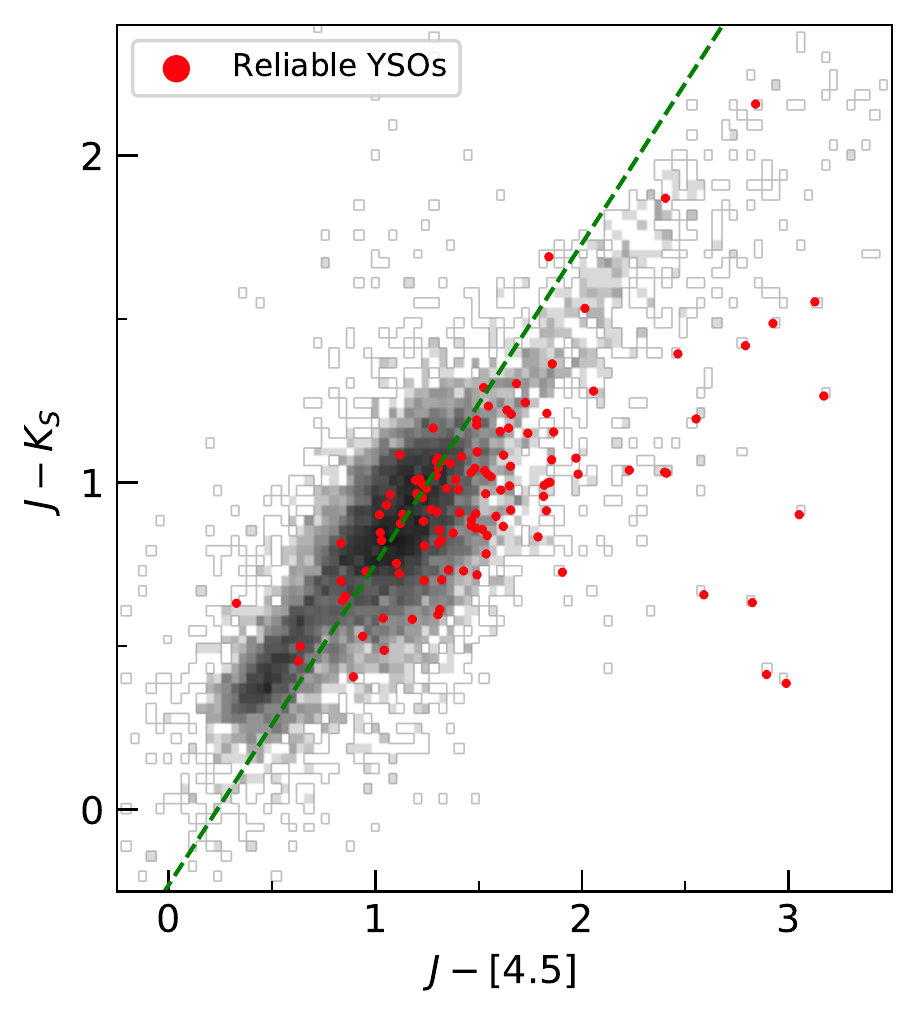}}
    \subfloat[\label{subfig:J_H_H_45_Hess}]{\includegraphics[width = 150 pt]{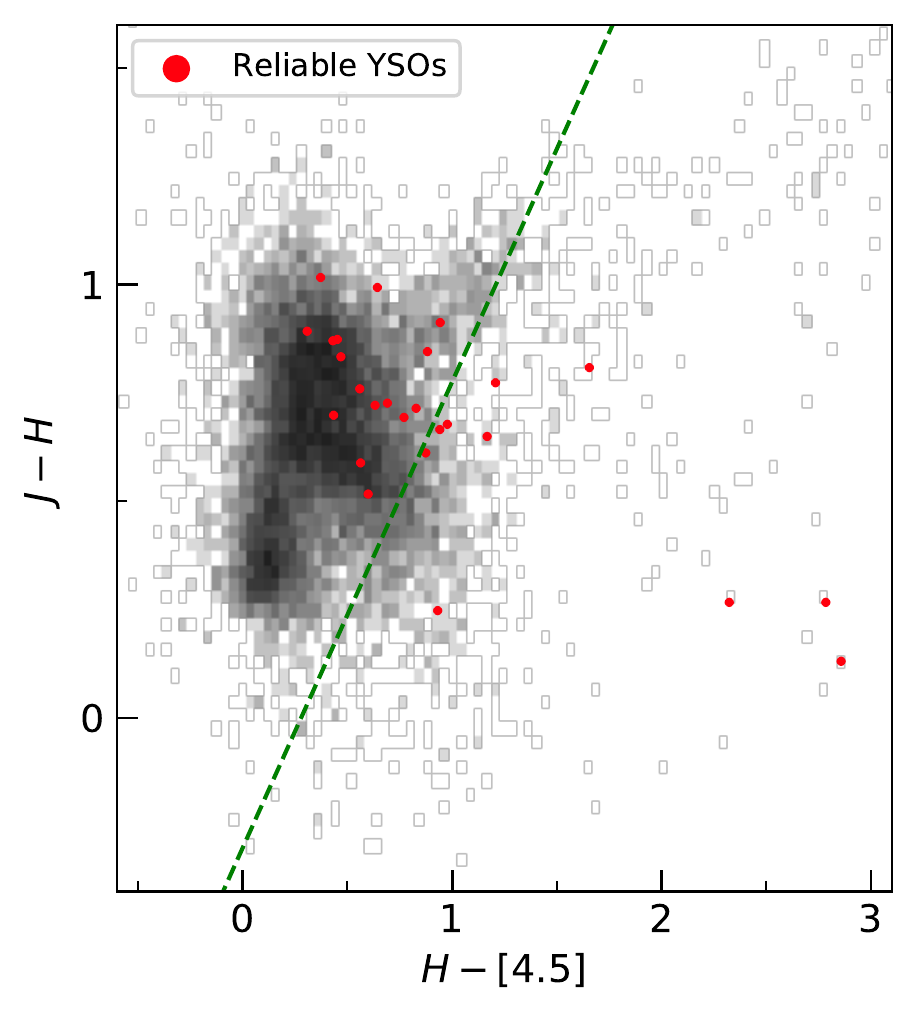}} \\
    \subfloat[\label{subfig:H_K_36_45_Hess}]{\includegraphics[width = 150 pt]{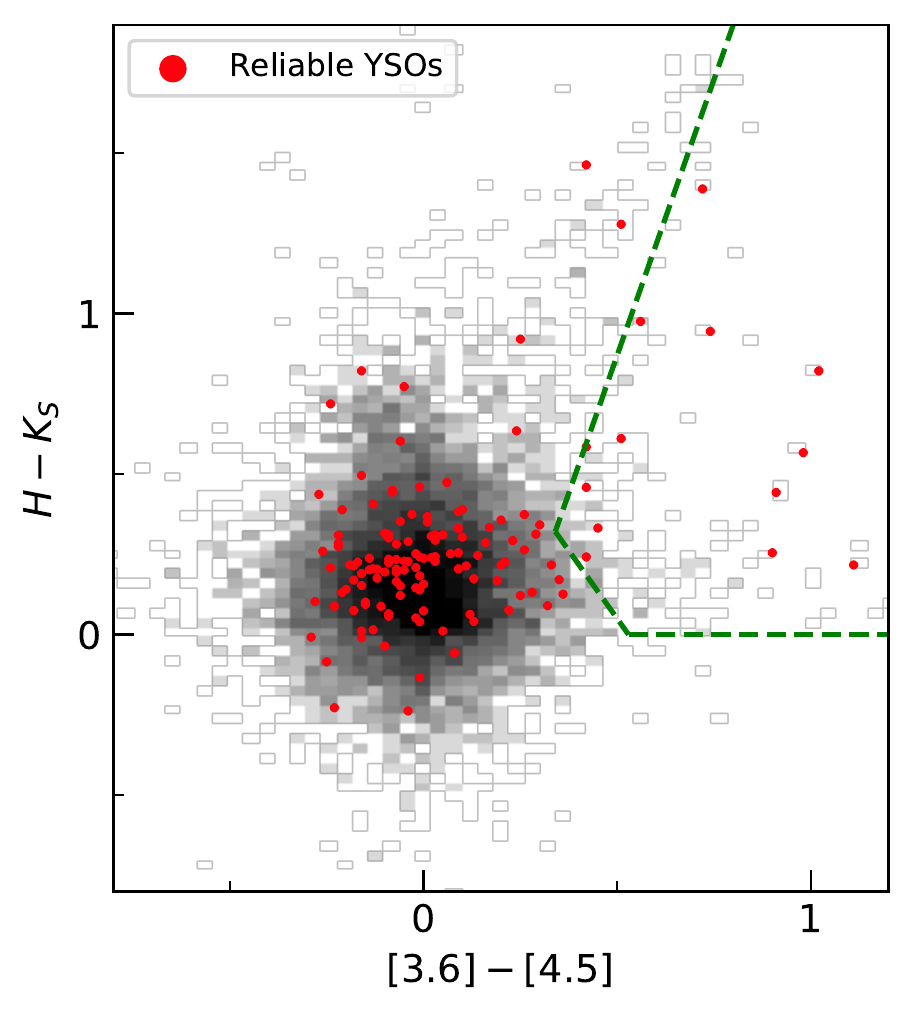}}
    \subfloat[\label{subfig:K_36_36_45_Hess}]{\includegraphics[width = 150 pt]{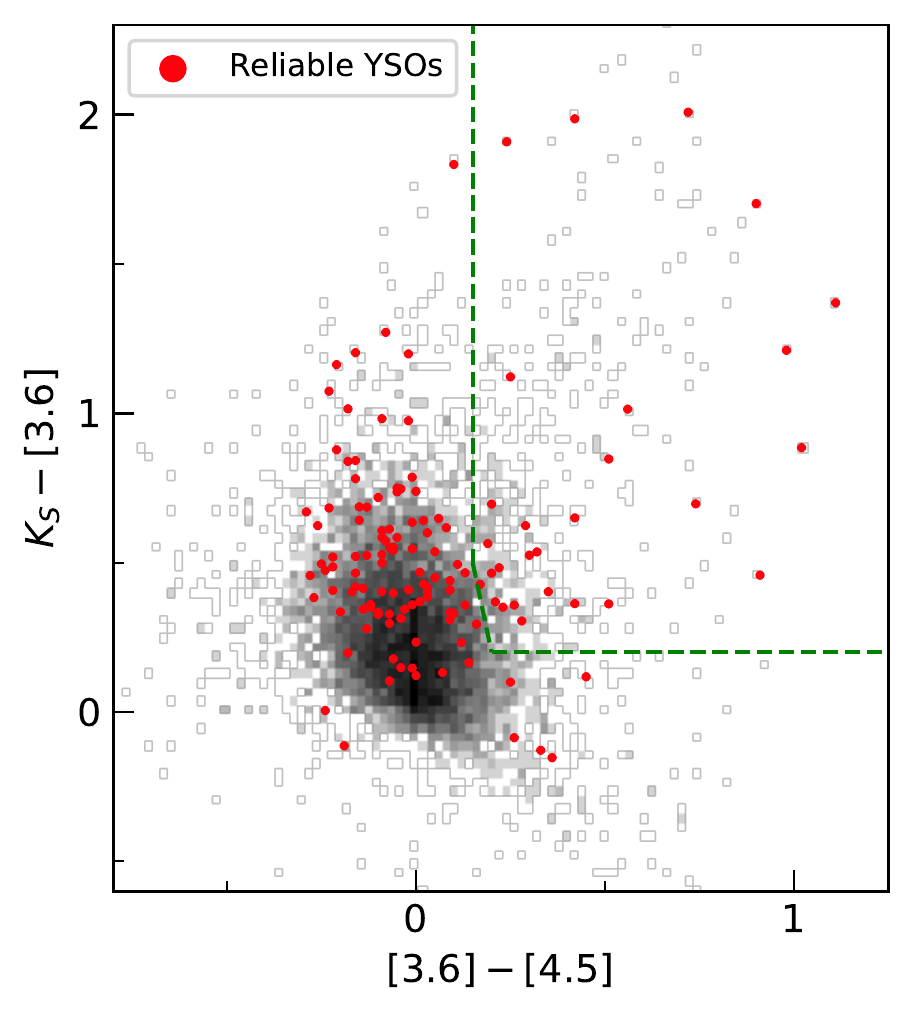}}
    \subfloat[\label{subfig:36_45_45_58_Hess}]{\includegraphics[width = 150 pt]{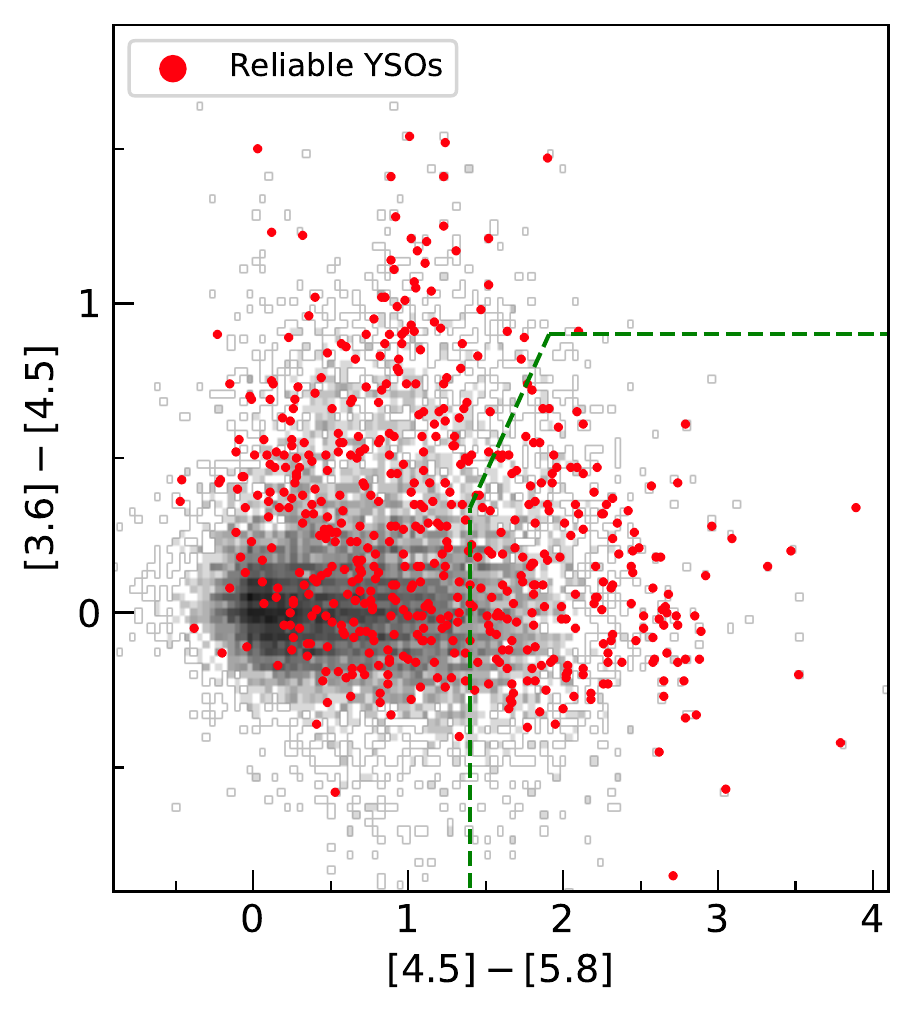}}
    \caption{Near-and mid-IR colour-colour diagrams for point sources in NGC 6822. The CCDs are represented as Hess diagrams. 
    The diagonal dashed line, in the top row, separates sources with an IR-excess (on the right) from those without. Sources enclosed by the box in the [3.6]--[4.5] vs.~[4.5]--[5.8] CCD have a significant level of PAH contamination.}
    \label{fig:CCDs}
\end{figure*}

In this analysis we include {\it Spitzer} sources that have no valid MIPS 24 $\mu$m photometry. Sources with low-quality photometry identified near the detection limits for the {\em Spitzer}  [5.8], [8.0] and [24] data may also be unreliable due to the distance. 
Instead, we use near-IR and IRAC 3.6 and 4.5 $\mu$m data to recover low signal-to-noise sources and retrieve candidate YSOs at an advanced stage of formation. 
To compensate for the relative drop in sensitivity in the red IRAC and MIPS bands at the distance of NGC 6822, we apply additional selection criteria to sources with at least two detections in $JHK$, that have IRAC 3.6 and 4.5 $\mu$m data with photometric uncertainties $\sigma < 0.2$ mag, and with no reliable [8.0] and [24] data. This ensures that we are not excluding likely YSOs in our previously unclassified objects. 
Sources with an IR excess were first identified using equation~\ref{eq:spitJHK}. A bright-source limit of $[3.6] \leqslant 16.5$ is then imposed to remove background galaxies.
Finally, we apply the colour cuts proposed by \citet{Gutermuth2009} and \citet{Koenig2014} to identify possible YSOs using only their $H-K$ vs.~$[3.6]-[4.5]$ and $K-[3.6]$ vs.~$[3.6]-[4.5]$ colours. 
The selection criteria is given in Eqs.~\ref{eq:HK_I1I2_cuts} and \ref{eq:KI1_I1I2_cuts}.  In total, we identified 131 YSO candidates which were missed by our mid-IR selection criteria.

\small
\renewcommand{\arraystretch}{0.75}
\begin{eqnarray}
\label{eq:spitJHK}
\left \lbrace
\begin{array}{r}
H-K <  1.1674 \cdot (K-[4.5]) - 0.5240 \; \;{\rm or} \\
 \\
  J-K < (J-[4.5]) - 0.4196 \;\;{\rm or}   \\
  \\
  J-H < 1.0758 \cdot (H-[4.5]) - 0.2997 \; \;\; \;\;\\
\end{array}     \right \rbrace 
\end{eqnarray}
\normalsize
\renewcommand{\arraystretch}{1}

\small %
\renewcommand{\arraystretch}{0.95}
\begin{eqnarray}
\label{eq:HK_I1I2_cuts}
\left \lbrace
\begin{array}{r}
H-K> 0.0 \; \;{\rm and} \; \;\;  H-K >  -1.7 \cdot ([3.6]-[4.5]) + 0.9 \\
{\rm and} \; \; H-K  < 3.44 \cdot ([3.6]-[4.5]) - 0.85 \\
\end{array}     \right \rbrace 
\end{eqnarray}
\normalsize
\renewcommand{\arraystretch}{1}

\small
\renewcommand{\arraystretch}{0.95}
\begin{eqnarray}
\label{eq:KI1_I1I2_cuts} 
\left \lbrace
\begin{array}{r}
~[3.6]  < 16.5   \; \; {\rm and} \; \;  K - [3.6]  > 0.5   \; \; \;\; \;\; \; \\
\; \; {\rm and} \; \; [3.6]-[4.5] \geqslant 0.15 \; \;{\rm or} \\
 \\
~[3.6]  < 16.5   \; \; {\rm and} \; \;  0.5 > K - [3.6]  > 0.2   \; \; {\rm and} \\
\; \;  K - [3.6]  > -6  \cdot  ([3.6]-[4.5]) + 1.4  \; \; \;\; \;\; \; \\
\end{array}     \right \rbrace 
\end{eqnarray}
\normalsize
\renewcommand{\arraystretch}{1}

The spatial distribution of the possible YSO candidates is more uniformly distributed than the high- and medium-confidence YSO candidates, which is indicative of contamination from background galaxies and field stars with ambiguous colours. 
To improve the reliability of candidate YSOs identified without a 5.6 $\mu$m, 8 $\mu$m or 24 $\mu$m flux, we require the possible YSOs to be spatially correlated with a star-forming region in NGC 6822, or an area of diffuse H$\alpha$ emission (see section~\ref{sec:spatialDist}).  As other nearby star-forming galaxies show that YSOs are not randomly distributed, but in fact are highly clustered near the sites of their formation \citep{Krumholz2018}. 
Clusters in the possible YSO candidate list were identified using the density-based spatial clustering of applications with noise (DBSCAN) algorithm \citep{Ester1996}. Points within a cluster can reasonably be regarded as a possible YSO candidate, while a noise point is probably a background galaxy, field star or other contaminant, and is excluded from our list. Table~\ref{tab:ngc6822_YSOtable_poss} lists the final selection of 584 possible YSO candidates in NGC 6822 star-forming regions, identified from the near-IR or with a mid-IR colour score of two. Candidates excluded from the YSO SED fitting on the basis of insufficient data points, disjointed SEDs, or the presence of a possible stellar photosphere (see Section~\ref{sec:contamination}) are also included in this category.

\subsection{Removing contaminants}
\label{sec:contamination}

Despite efforts to eliminate contamination from other populations using our extensive colour selections, some level of contamination is to be expected in the YSO sample. 
At the distance of NGC 6822, unresolved background galaxies, including active galactic nuclei (AGN) and star-forming galaxies, have mid-IR colours that overlap significantly with YSOs. Magnitude-based cutoffs which are very effective at selecting YSOs in the Magellanic Clouds are less efficient in more distant galaxies like NGC 6822 due to the lower luminosity of the population. To estimate a conservative upper limit for galaxy contamination, we assume that unresolved galaxies are faint and evenly distributed across the field.  We select a 4.32 arcmin$^2$ area on the outskirts of the galaxy away from known star-formation regions to estimate the approximate level of background contamination in our YSO cuts. This region has a total point-source density of 37.04 sources/arcmin$^2$ and a reliable YSO candidate density of 0.23 sources/arcmin$^2$, which corresponds to a $\sim$15\% contamination level in our global YSO candidate list.  The estimated contamination level for each colour selection is given in Table~\ref{tab:ngc6822_maxContamLevels}. As this region still contains NGC 6822 member sources and is within the H\,{\sc i} disk, however, it is possible that we have overestimated the extent of this extragalactic contamination.

\begin{table}
\centering
\begin{minipage}{84mm}
 \caption{Contamination estimates in each colour selection criteria used to select YSOs for further analysis.}
 \label{tab:ngc6822_maxContamLevels}    
\centering
 \begin{tabular}{lcc}
   \hline
   \hline
   CMD   &  Number of colour  & Max. contamination \\
         & selected YSOs    & (percent) \\
\hline   
   $[3.6]$ vs. [3.6]--[5.8]  &   787    &       49.2       \\      
   $[3.6]$ vs. [3.6]--[8.0]  &   639    &       15.2       \\      
   $[4.5]$ vs. [4.5]--[5.8]  &   875    &       55.4       \\      
   $[8.0]$ vs. [4.5]--[8.0]  &   897    &       21.6       \\      
   $[4.5]$ vs. [4.5]--[24]   &   911    &       10.6       \\      
   $[8.0]$ vs. [8.0]--[24]   &   654    &       14.8       \\      
   Reliable catalogue        &  584     &      16.6       \\    
  \hline
 \end{tabular}
\end{minipage}
\end{table}

YSOs are typically found in regions with emission from PAH molecules and dust grains, which can contaminate the {\em Spitzer} fluxes. 
Diffuse emission from PAHs affects the 3.6, 5.8, and 8.0 $\mu$m {\em Spitzer} IRAC bands. 
The [3.6]--[4.5] vs. [4.5]--[5.8] colour-colour diagram (CCD) shown in Figure~\ref{fig:CCDs} can be used to assess contamination from PAH emission in our sample \citep[e.g.][]{Carlson2012}. Sources in the presence of the strong PAH emission occupy a distinct area in the CCD which extends towards the lower-right and are outlined by a dashed line in Figure~\ref{fig:CCDs}.  We note these PAH-enhanced sources in Tables~\ref{tab:ngc6822_HP_YSOtable}, \ref{tab:ngc6822_YSOtable_prob}, and \ref{tab:ngc6822_YSOtable_poss}.
%
Conversely, no known globular clusters display active star formation. 
There are three known globular clusters in our FOV, Hubble VII, SC3, and SC6 \citep{Veljanoski2015}, listed in Table~\ref{tab:ngc6822_GCs}. We identify two YSO candidates within their half-light radii and remove them from our sample. 

\begin{table}
\centering
\begin{minipage}{84mm}
 \caption{Globular Clusters in the {\em Spitzer} field.}
 \label{tab:ngc6822_GCs}    
\centering
 \begin{tabular}{lccc}
   \hline
   \hline
   ID   &  RA & DEC & $r_{\rm h}$ (pc) \\
\hline   
Hubble-VII 	& 19 44 55.8  &	-14 48 56.2 & 2.5 \\
SC3         & 19 45 40.2  & -14 49 25.8 & 7.5 \\
SC6         & 19 45 37.0  & -14 41 10.8 & $\dots$ \\
  \hline
 \end{tabular}
\end{minipage}
\end{table}


Resolution limitations of {\em Spitzer} can confuse the colour selection.
At the distance of NGC 6822, source confusion and flux enhancements in our catalogues are a limitation of our data. Star-forming regions are inherently crowded, and our catalogue is both crowding- and magnitude-limited.  The aperture diameter of MIPS (6$^{\prime\prime}$ corresponds to  $\sim14.4$ pc at our adopted distance) is considerably larger than the FWHM of the IRAC 3.6 $\mu$m PSF ($1.7^{\prime\prime}, \sim 4$pc), which is larger still compared to the $JHK$ data. Rather than isolated sources, a single YSO data point is likely to contain multiple objects and even entire star-formation clusters that make up a local surface density enhancement. These {\em Spitzer} fluxes are dominated by the most massive object in the protoclusters, owing to the steep mass-luminosity relation \citep{Seale2009, Oliveira2013}. 
Multiple sources within a single data point likely belong to the same cluster, however it is also possible that other emission may also contaminate the beam resulting in a flux enhancement. 


Sources with spectral energy distributions (SEDs) that rise from 8 to 24 $\mu$m are typically YSOs, compact \Hii~regions, unresolved background galaxies, or planetary nebulae (PNe). Visual inspection of the YSO candidates' SEDs reveal a significant number of sources with both a stellar photosphere component and an infrared excess component not directly associated with the point-source. This can be explained by the limited resolution of the mid-IR imaging detecting stars embedded within larger ISM structures in the longer-wavelength images or multiple sources in the beam. 
This can be especially problematic in star-formation regions. 

\begin{figure}
    \raggedleft
    \includegraphics[trim=5 35 1 1,clip, width=0.5\textwidth]{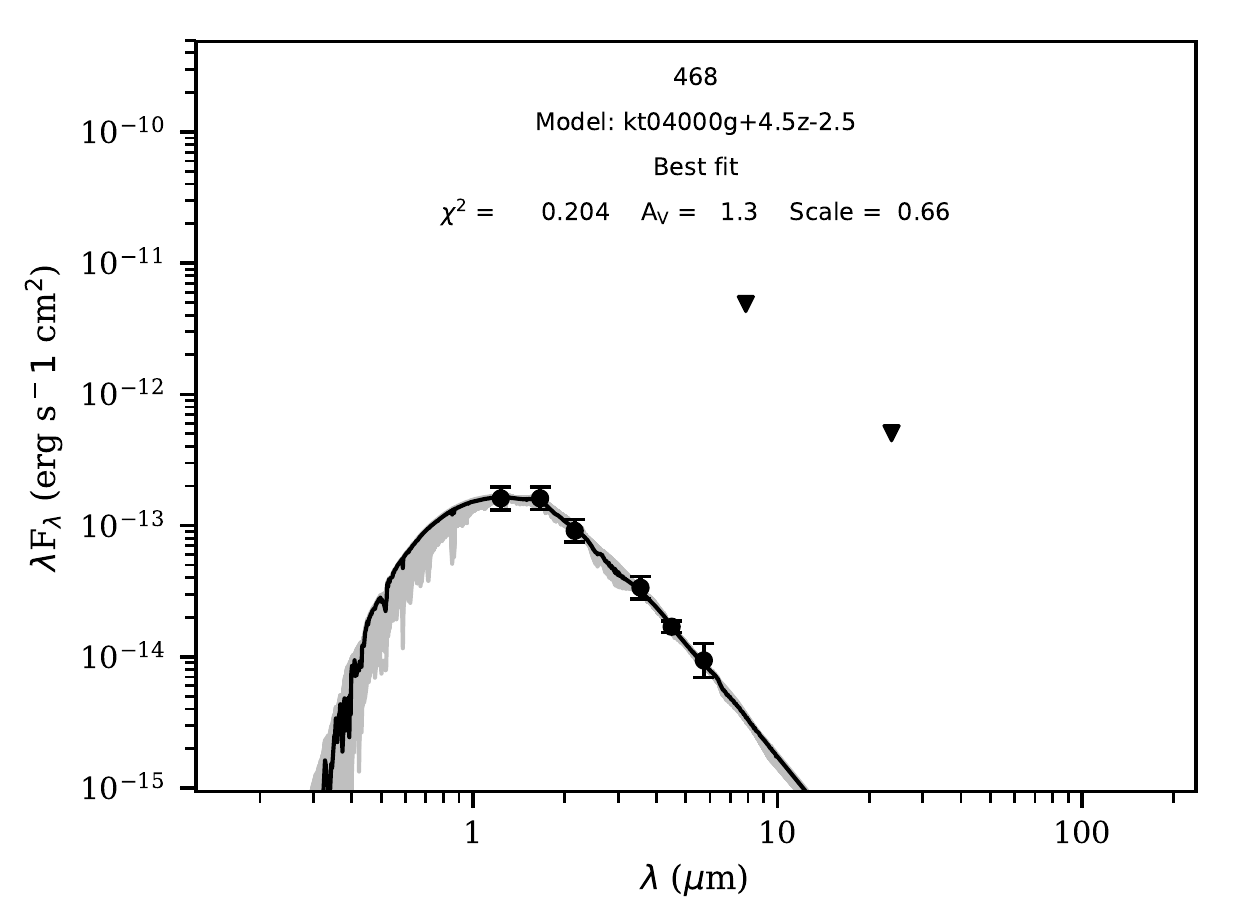}
    \includegraphics[trim=5 35 1 1,clip, width=0.5\textwidth]{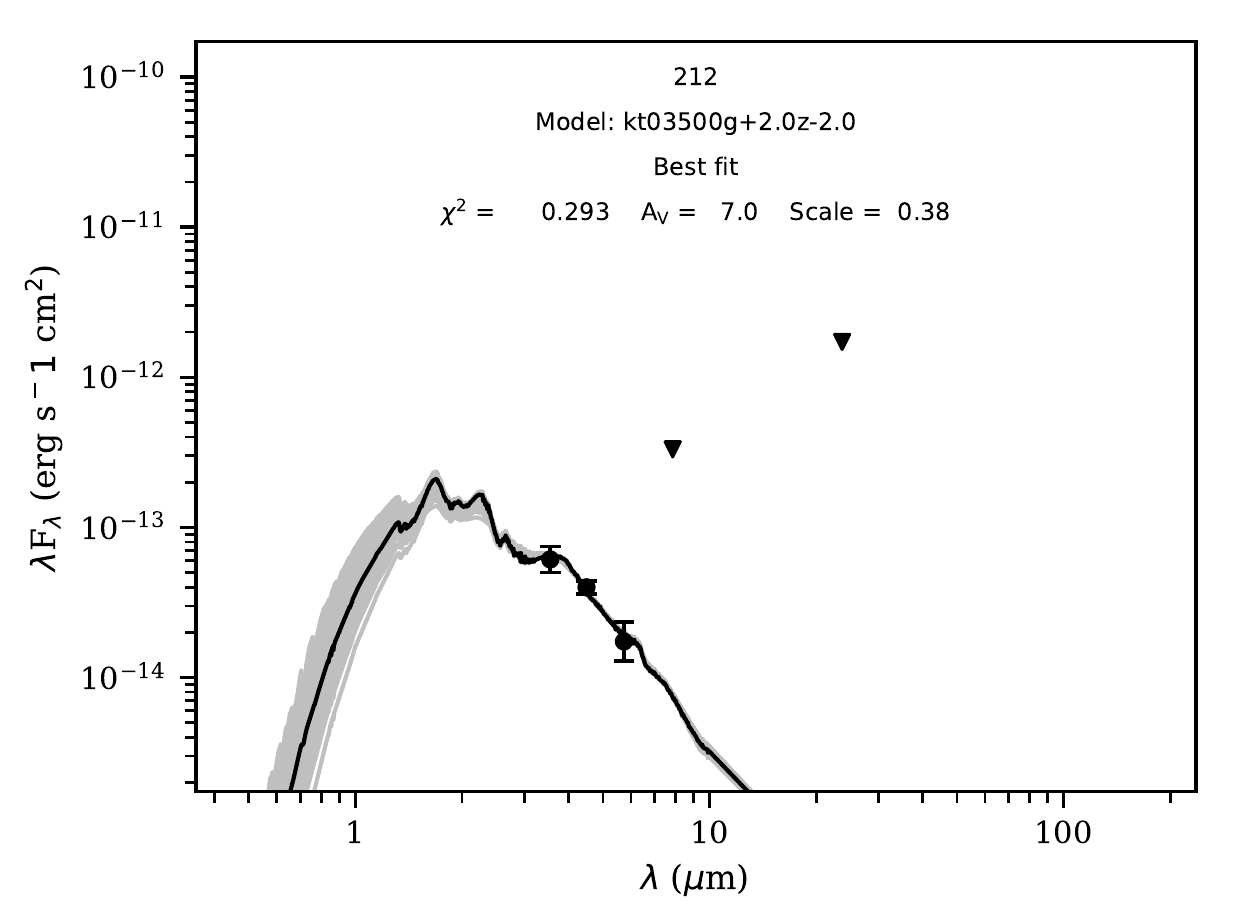}
    \includegraphics[trim=5 1 1 1,clip, width=0.5\textwidth]{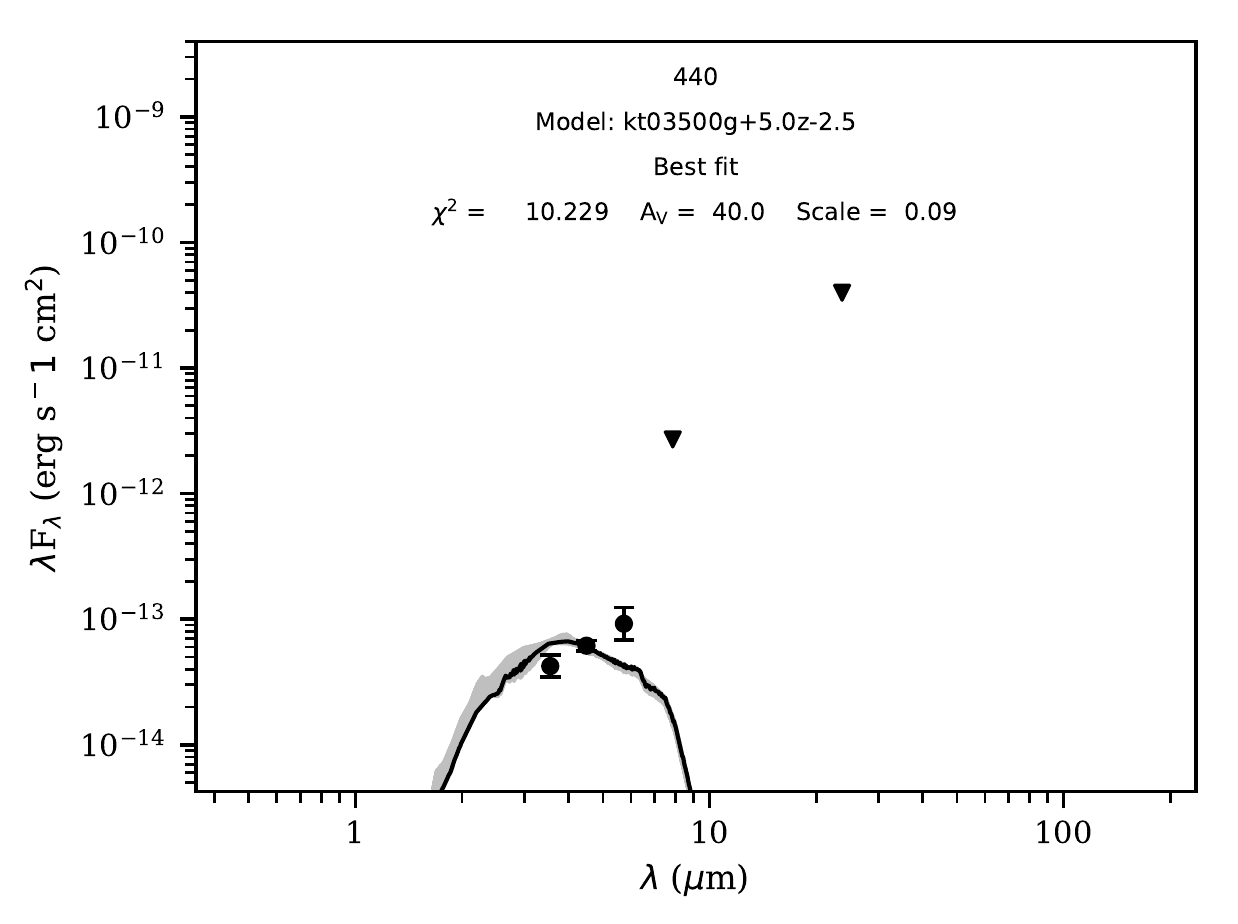}
    \caption{SEDs of candidate YSOs fit with the~\protect\cite{Castelli2003} grid of stellar atmospheres. The top two panels show the SEDs (black circles) and their associated best-fit model spectra (black line) of two sources which are well-fitted by a photosphere. The bottom panel shows an example SED not well represented by a stellar photosphere. Only the last sources is retained in our `probable' YSO candidate list. Black triangles denote upper-limits to the photometric data points used in the SED fitting.}
    \label{fig:SED_photosphere}
\end{figure}

To ensure only photometry associated with one astronomical object is included in our analysis of each YSO SED and our cross-matched data belongs to a red point-source or compact cluster, we fit SEDs of the YSO candidates with a grid of Kurucz {\sc atlas9}  model atmospheres from \cite{Castelli2003}, using the \cite{Robitaille2007} SED fitting tool (see Section~\ref{sec:sedfit}), and remove sources that are well fitted by a stellar photosphere.
%
Before SED fitting, we inspect each individual SED and discarded sources with the following criteria: too few valid data points ($<$5 bands) in the combined catalogue; disjointed SEDs due to a mismatch between catalogues; sources with fewer than  three {\em Spitzer} detections; and sources without an 8 or 24 $\mu$m detection, such that the fitting can be better constrained.  518 sources (89\%) of the colour-selected YSO sample meet these criteria. 
During the fitting procedure for the stellar photospheres, both the 8 and 24 $\mu$m fluxes were treated as upper limits, to account for possible enhancements from PAH emission and warm dust in the large MIPS beam, respectively. Figure~\ref{fig:SED_photosphere} shows example fits to the candidate YSO SEDs by the model photospheres. In total, we have removed 305 YSO candidates during this process, and flagged a further 20 as potential contaminants, leaving 213 sources.

Finally, we searched the SIMBAD and VizieR Astronomical Databases using the CDS X-Match Service\footnote{\url{http://cdsxmatch.u-strasbg.fr/##tab=xmatch}}  
to check if any of the YSO candidates are associated with a known astronomical object. 89 of the high confidence YSO candidates were matched within a 2$\arcsec$ radius. 
Of these sources, five have a spectroscopic classification; 73 had a literature classification indicative of youth, e.g., H$\alpha$-emitting objects, CO-bright clumps, compact H\,{\sc ii} source, or location in a star-formation region \citep{Massey2007, Schruba2017, HernandezMartinez2009, Melena2009}; three had multiple conflicting classifications; whilst the remaining seven sources were matched to a candidate AGB star, identified from optical or near-IR colours \citep{Sibbons2012, Whitelock2013, Kang2006, Letarte2002}. In total, $<$10 \% of our colour selected YSO candidates have a conflicting classification to our own. 

Evolved stars can have significant IR emission due to circumstellar dust.  In general, AGB stars are more luminous than YSOs and have colours which are not quite as red. Only the most extreme AGB stars which are surrounded by a significant amount of dust, for instance OH/IR stars, have colours similar to YSOs in the mid-IR. In metal-poor galaxies like NGC 6822, these objects are rare \citep{Jones2015b, Jones2018} and thus are unlikely to pose a significant level of contamination to our YSO lists. 
We do not exclude any source with a literature classification which is based solely on colour-magnitude cuts, as YSOs and AGBs are easily confused in optical and near-IR colour space. 

 Rather surprisingly, a spectroscopically confirmed Red Giant \citep{Kirby2017} is matched to one of our YSO candidates. This is probably a result of a superposition of sources; one optically bright and one IR bright in this crowded field. 
 
Unlike AGB stars, unresolved planetary nebulae (PNe) are potentially a source of significant contamination in our YSO sample.
They have mid-IR {\em Spitzer} colours which are indistinguishable from YSOs and spectroscopic data are needed to segregate these objects \citep{Jones2017b}. 
There are currently 26 known PN candidates in NGC 6822, most of them located along the optical bar \citet{HernandezMartinez2009}. Indeed, our literature search finds that four spectroscopically confirmed planetary-nebular and three PNe candidates identified via their [O\,{\sc iii}] $\lambda$5007/H$\alpha$ flux ratio \citep{HernandezMartinez2009, GarciaRojas2016} are included in our YSO candidate list.  We exclude them from further analysis. 

\section{Radiative transfer modelling of YSO candidates}
\label{sec:Models}

YSOs can be classified according to their evolutionary stage based on the physical properties of their circumstellar dust distribution \citep{Robitaille2006} rather than the observational characteristics of its SED \citep{Lada1987}. 
To determine the evolutionary stage and physical properties of our YSO candidates, we fit their SED with the grid of radiative transfer models developed by \citet{Robitaille2006, Robitaille2017} hereafter \citetalias{Robitaille2006, Robitaille2017} respectively. 

\subsection{YSO model grids}

\subsubsection{Robitaille et al.~(2006) Model Grid}

The axisymmetric YSO model grid of \citetalias{Robitaille2006} spans a range of different stellar, disk, and envelope properties at 10 inclination angles, resulting in a set of 200,000 model SEDs. These models were computed using the \citet{Whitney2003a, Whitney2003b} radiative transfer code for a central stellar mass between 0.1 -- 50 ${\rm M}_{\odot}$ with an age between 10$^3$ -- 10$^7$ yr. The pre-main-sequence stars have varying combinations of circumstellar geometry including a flared accretion disk and an infalling, rotationally-flattened envelope which can include cavities carved out by a bipolar outflow.  
In total, 14 stellar and circumstellar dust geometry parameters are varied to produce the model grid. These include: stellar luminosity, temperature, disk mass, radius, inner radius, envelope infall rate, and bipolar cavity opening angle. Thus these YSO models include the youngest, most embedded YSOs at the early stage of envelope infall to the late disk-only stage and ultimately pre-main-sequence stars with little or no dusty disk.
Fits with the \citetalias{Robitaille2006} models can then be directly compared to star-formation results for the LMC and SMC \citep{Whitney2008, Chen2010, Carlson2012, Sewilo2013}.

\subsubsection{Robitaille et al.~(2017) Model Grid}

The \citetalias{Robitaille2017} models are a set of eighteen different YSO model grids of increasing complexity. 
In these models the central protostellar source may be associated with a disk, a circumstellar infalling envelope, bipolar cavities, and an ambient medium.
Each SED was computed for nine viewing angles using a random stratified sampling between 0$^{\circ}$ and 90$^{\circ}$. The dust properties are the same for all models and do not include polycyclic aromatic hydrocarbon (PAH) emission. 
The simplest model has two free parameters: a stellar radius $R_{\star}$ and an effective temperature $T_{\rm eff}$, which defines the central irradiating source.
Conversely, the most complex set of models, which includes all possible physical components, has twelve free parameters. These parameters have been uniformly sampled and do not introduce correlations, at the cost of some parameter combinations being nonphysical. For more details on each component in the models and their free input parameters, we refer the reader to \citetalias{Robitaille2017}.

\subsection{SED Fitting procedure}
\label{sec:sedfit}

We use a Python-based SED-fitting tool\footnotemark \footnotetext{https://sedfitter.readthedocs.io/en/stable/index.html} \citep{Robitaille2007} to fit the  large set of YSO models to the photometric data, as this accounts for errors in the distance to the source, the effects of interstellar extinction ($A_{\rm V}$), and aperture size. As before we assume a distance to NGC 6822 of 490 kpc with a error of $\pm$8\%. 
Prior to fitting, photometric uncertainties in each band were adjusted so that a lower limit of 10\% on the observed flux was introduced, in order to account for variability \citep[e.g.][]{MoralesCalderon2011, Gunther2014} and other systematic and calibration errors in the data. The models were fitted to the near-IR and {\em Spitzer} photometry of all our YSO candidates listed in Tables~\ref{tab:ngc6822_HP_YSOtable} and~\ref{tab:ngc6822_YSOtable_prob}, providing they meet the photometric-quality criteria listed above.

PAH emission is not accounted for in the models. Sources contaminated by PAHs along the line of sight have a distinctive dip at 4.5$\mu$m in their SEDs, the intensity of which depends on the strength of the PAH emission.
In total, 117 SEDs have IRAC data which is clearly affected by emission from PAHs; only the [4.5] band is unaffected \citep{Churchwell2004}, and should be assigned a higher weight in the fitting.   
To account for this we adopt the PAH correction method employed by \cite{Carlson2012} and increase the error bars to 20\%, 10\%, 30\%, and 40\% in the 3.6, 4.5, 5.8, and 8.0 $\mu$m bands to fit sources with significant PAH contributions.


\begin{figure*}
    \centering
    \includegraphics[width = 0.49\textwidth]{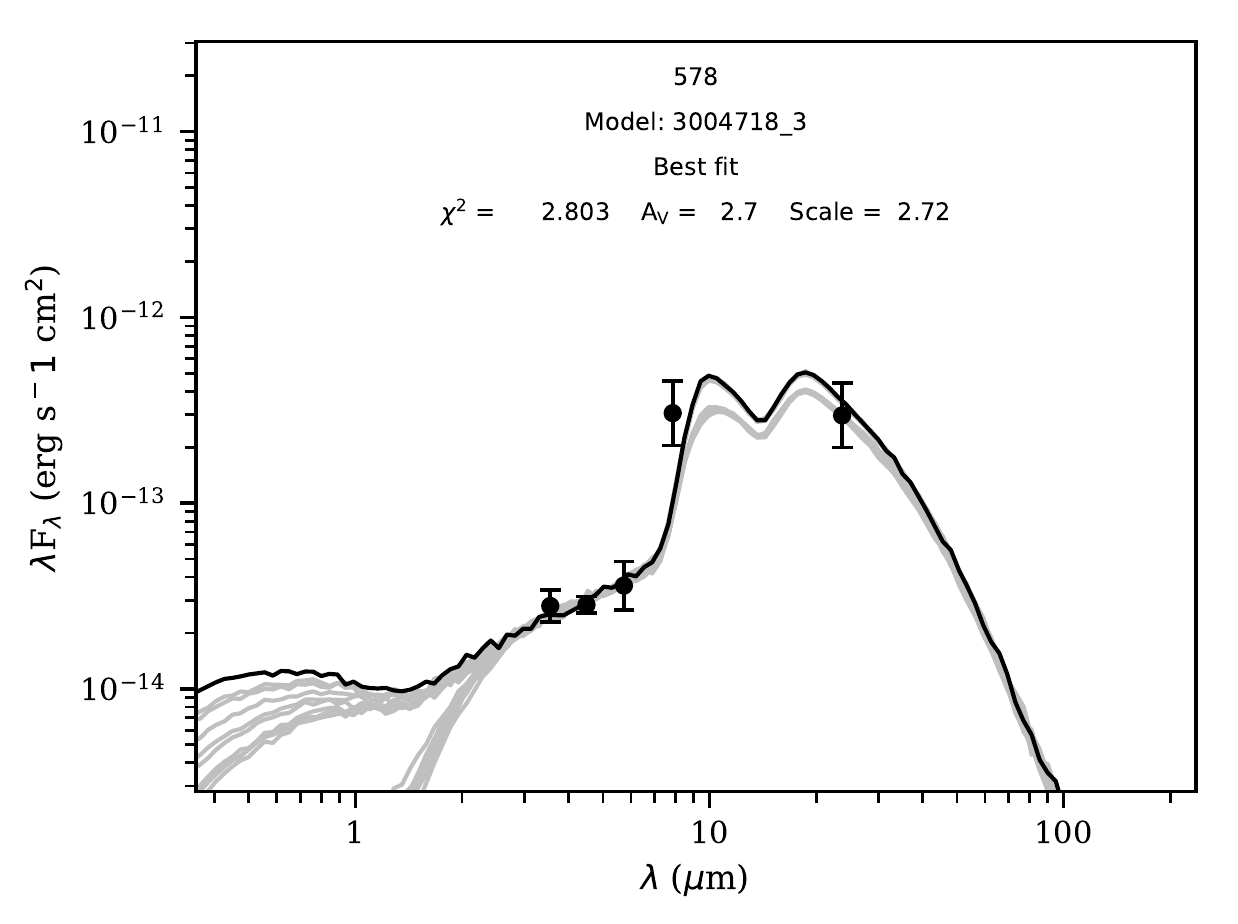}
    \includegraphics[width = 0.49\textwidth]{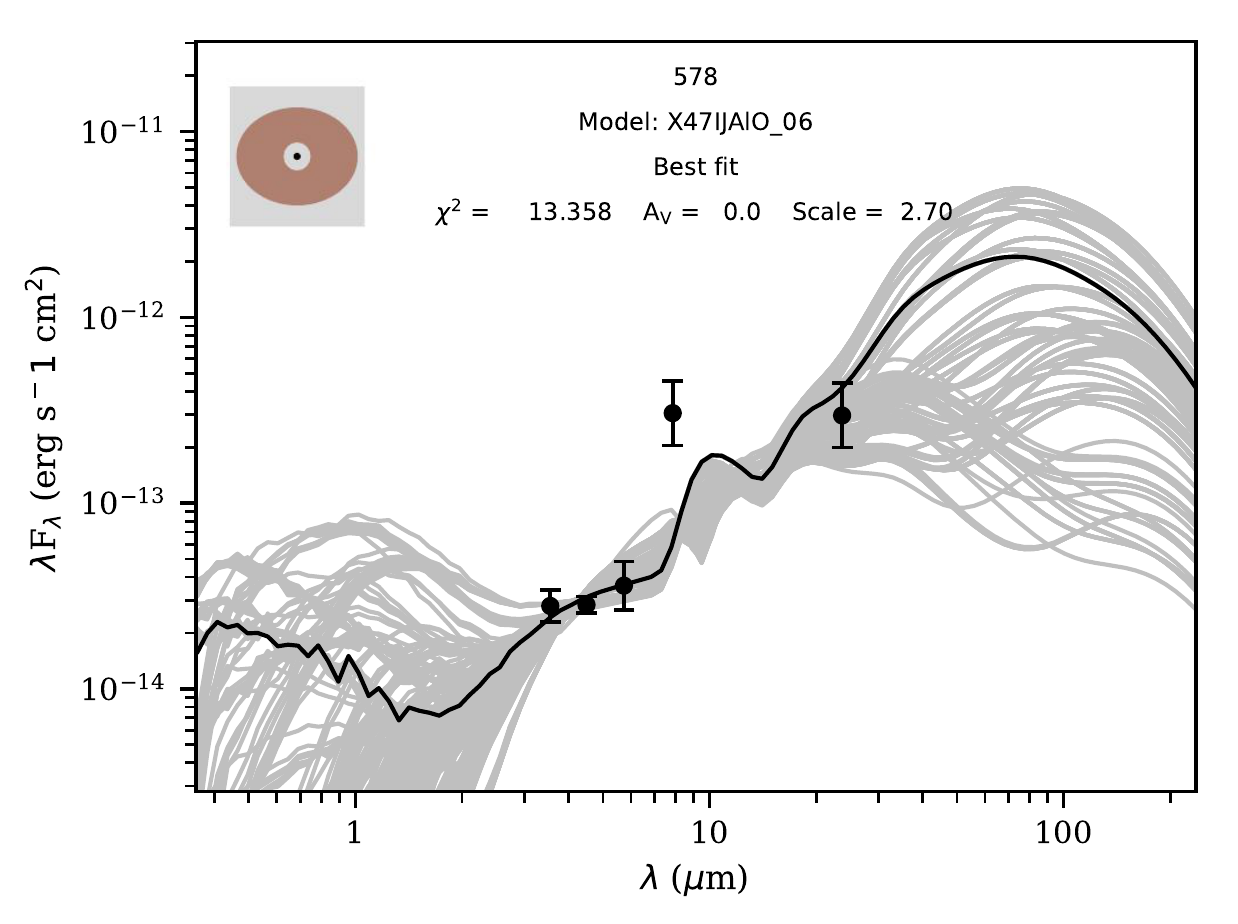}    
    \includegraphics[width = 0.49\textwidth]{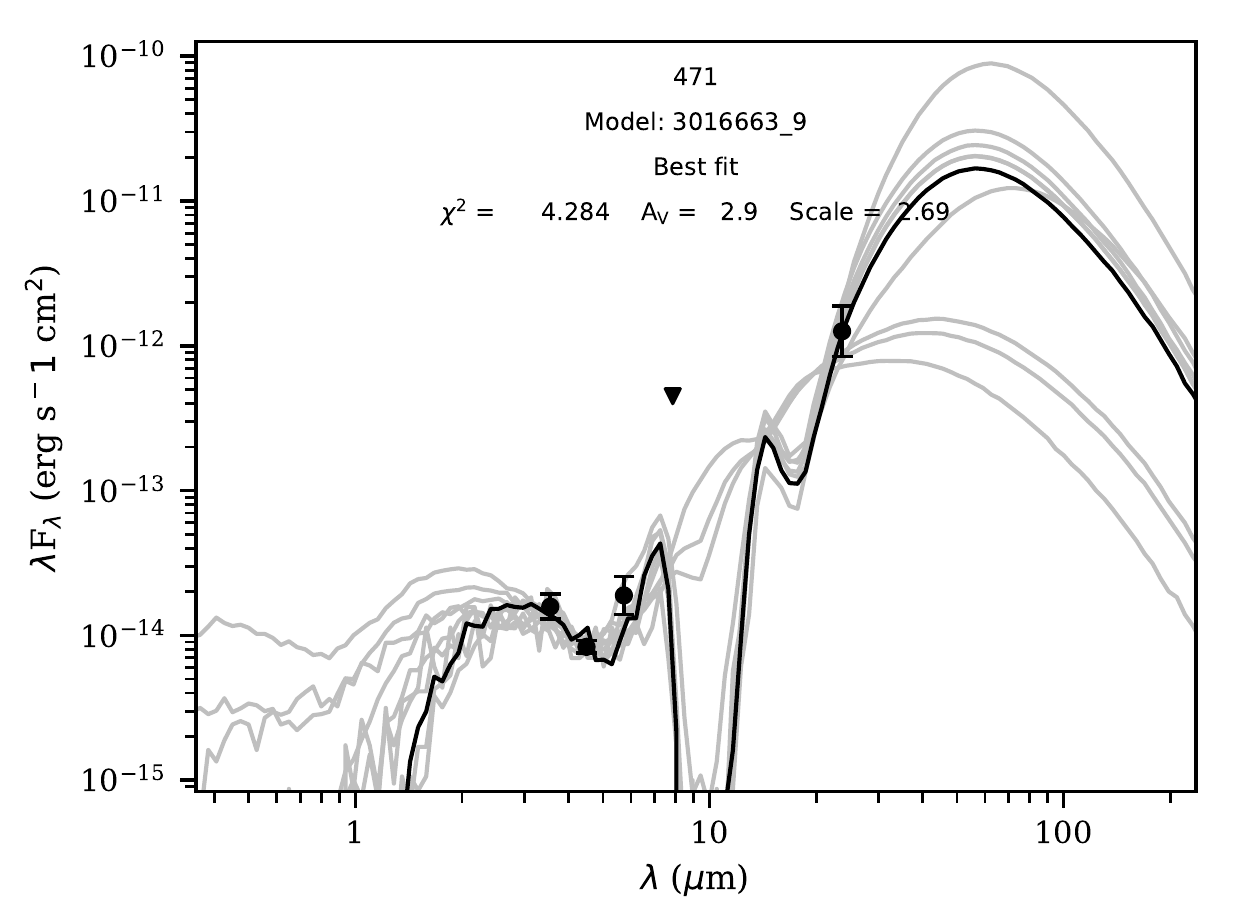}
    \includegraphics[width = 0.49\textwidth]{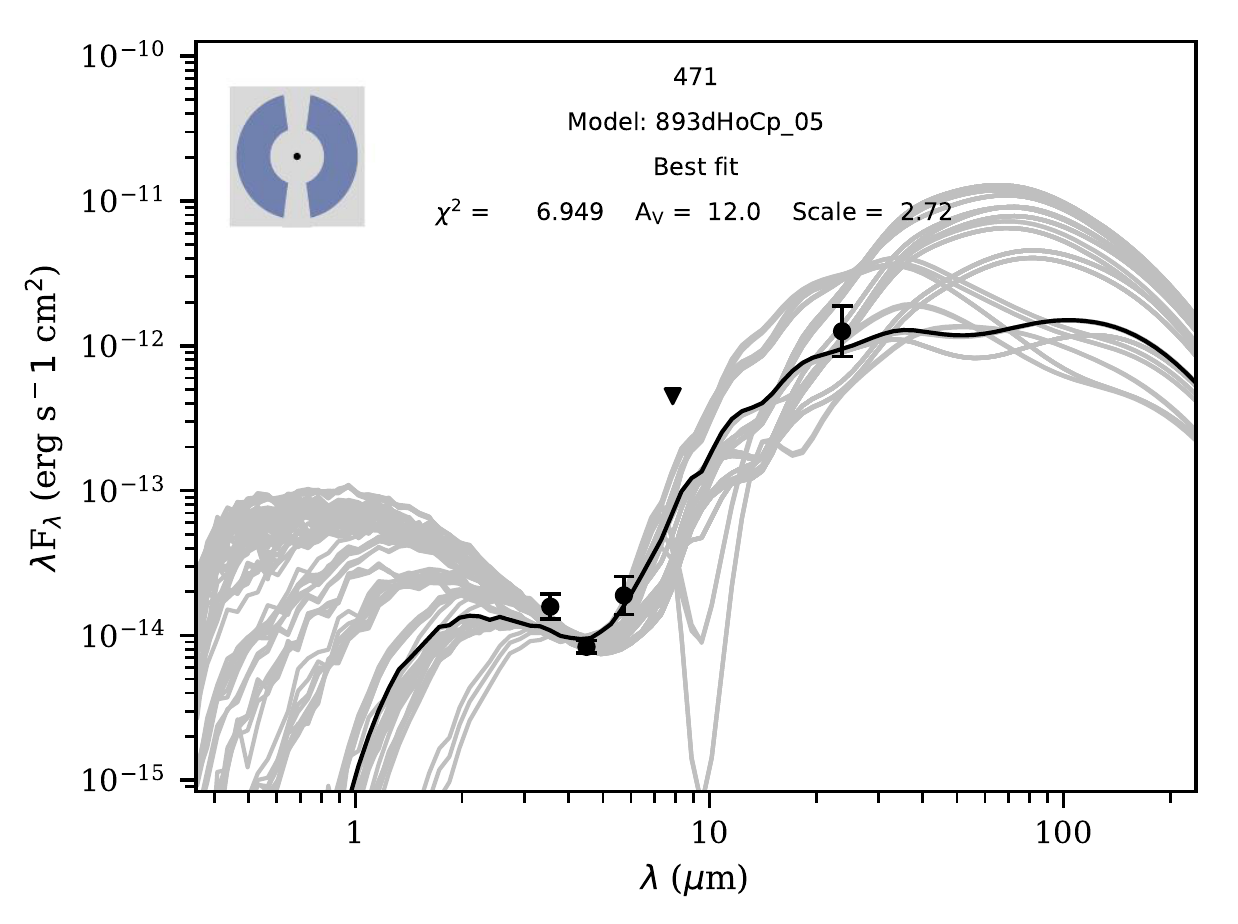}    
    \includegraphics[width = 0.49\textwidth]{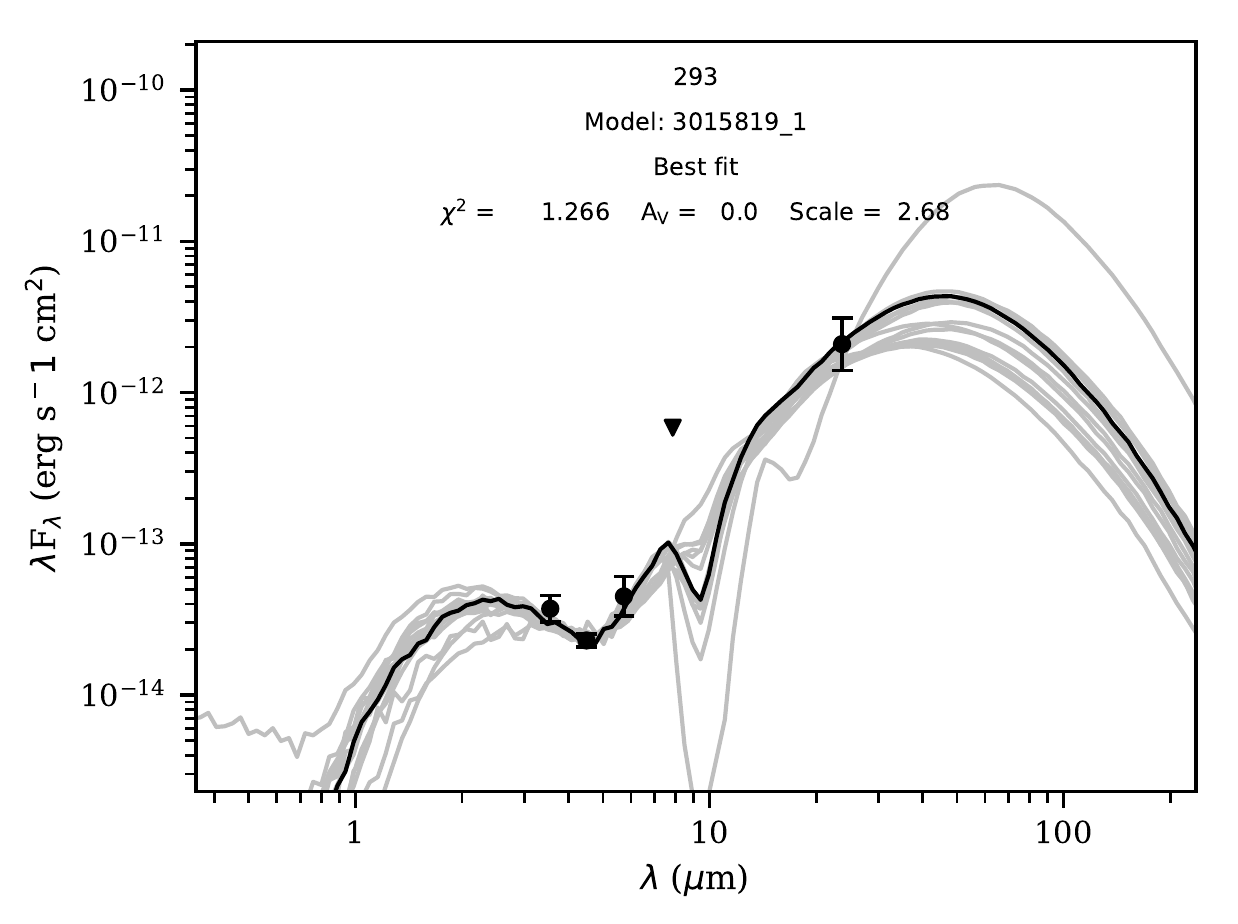}
    \includegraphics[width = 0.49\textwidth]{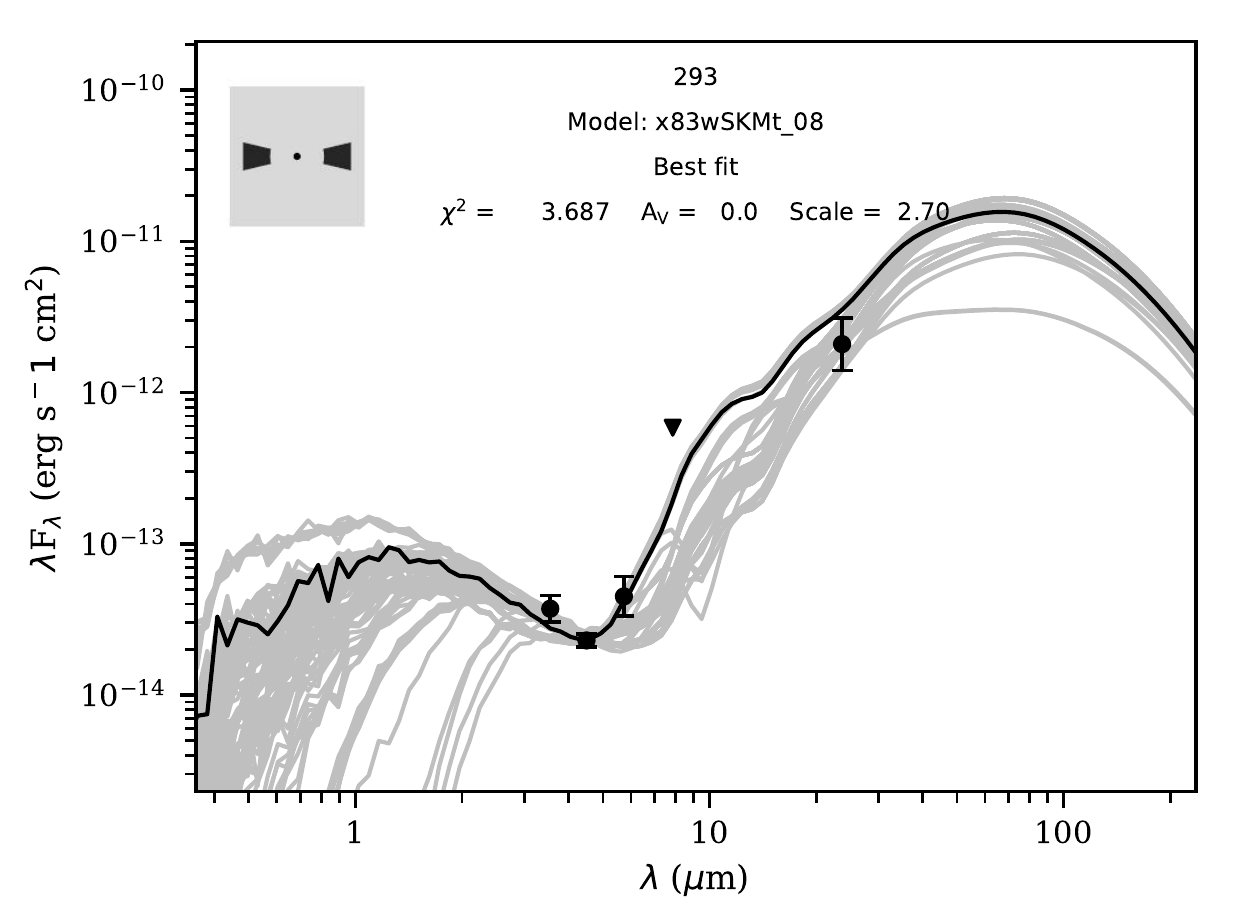}
    \caption{Example SEDs of the 'high-confidence' YSOs fit with the~\protect\citetalias{Robitaille2006} (left) and~\protect\citet{Robitaille2017} models (right). The observed SED and photometic errors are shown as filled circles; upper limits are plotted as black triangles. The best-fit YSO model is shown as a solid line, grey lines indicate the range of alternate models with acceptable fits to the data (defined by $\chi_{\nu}^{2} - \chi^{2}_{\rm best} < 3 \times n_{\rm data}$). The source ID, best-fit model number, $\chi^2$, foreground extinction, and scale factor are shown in the label. The geometry of the best model-set for the~\protect\citetalias{Robitaille2017} models is also indicated.} 
    \label{fig:SEDmodelfit}
\end{figure*}


\begin{figure*}
    \centering
    \includegraphics[width=0.49\textwidth]{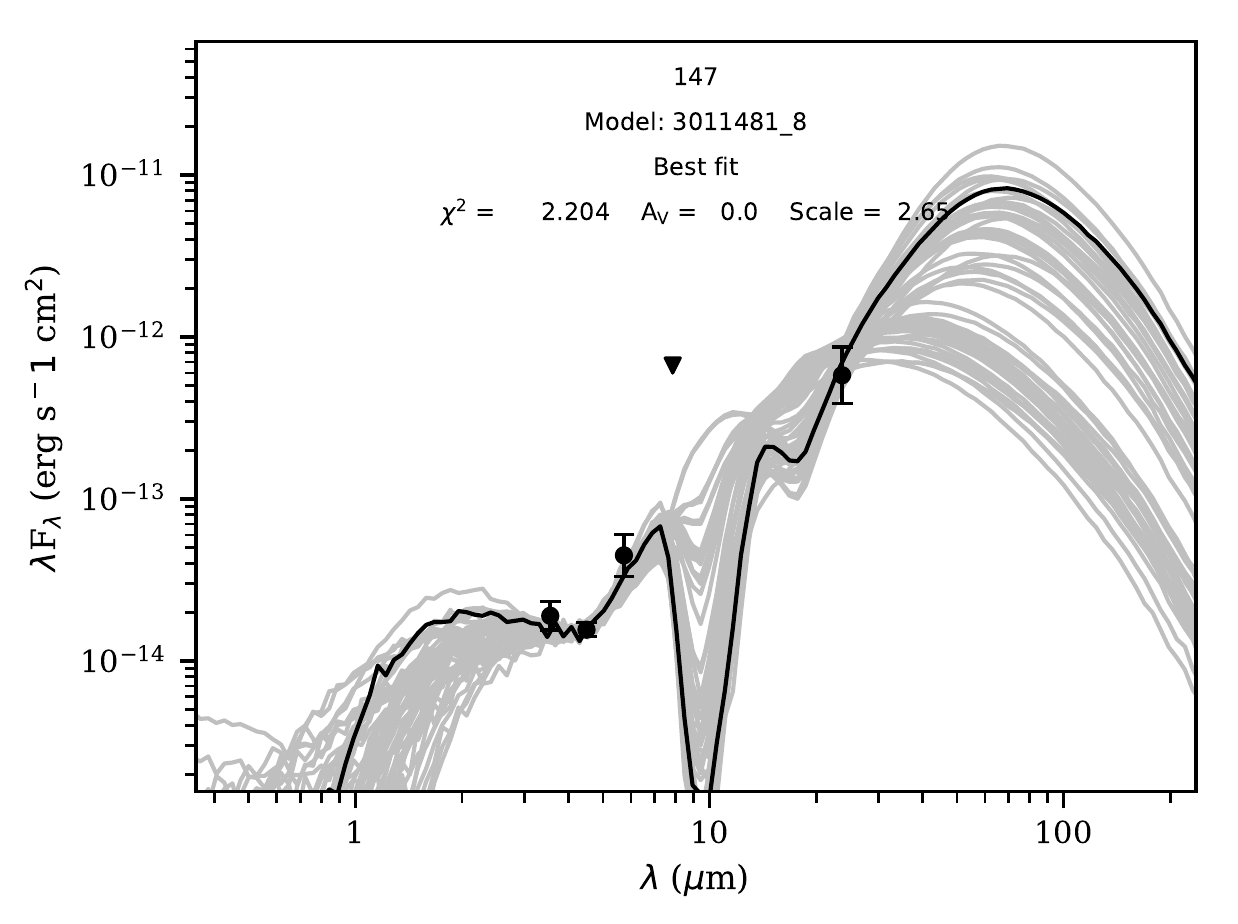}
    \includegraphics[width=0.49\textwidth]{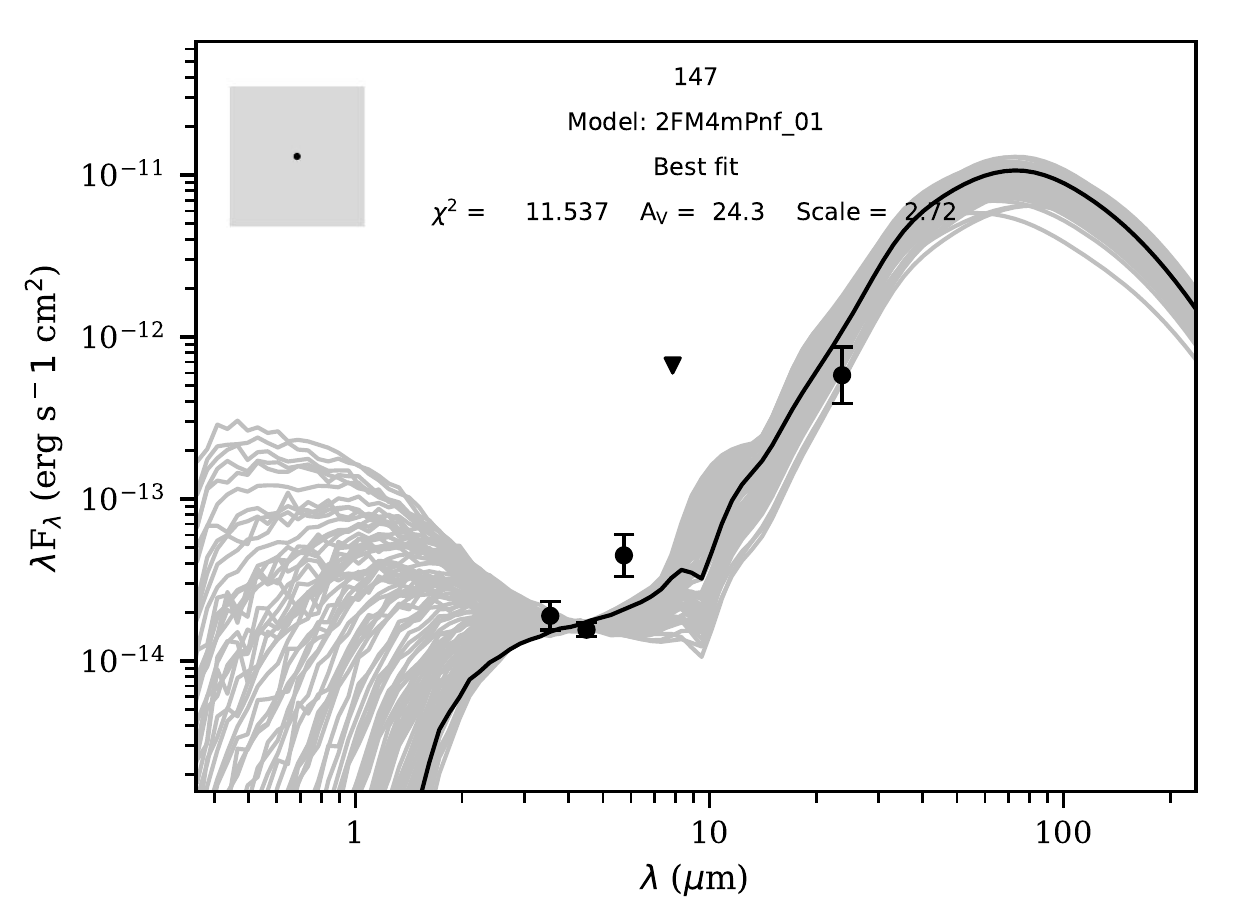}
    \caption{Source 147 best fit by as Stage I YSO by the ~\protect\citetalias{Robitaille2006} models, compared with its \protect\citetalias{Robitaille2017} best-model set consisting of a star and an ambient ISM.  Symbols as in Figure~\ref{fig:SEDmodelfit}. } 
    \label{fig:SEDmodelfit_17v06}
\end{figure*}

55\% of the candidate YSOs were satisfactorily fit by the YSO models. We consider these sources to be `high-confidence' YSO candidates. Figure~\ref{fig:SEDmodelfit} displays the SEDs and model fits to some example sources selected to illustrate the range of YSO parameters spanned by our sample at different evolutionary stages. 
Given the model complexity and the number of input parameters, there is degeneracy in the model output. 
Several models can produce a valid and almost equally good fit to the data; to prevent over-fitting, we consider the YSO models with the lowest $\chi_{\nu}^{2}$ values and use these to estimate the median absolute deviation uncertainty in the best-fit model parameters. 
For each source and model set, we identify the model with the lowest reduced $\chi_{\nu}^{2}$, namely $\chi^{2}_{\rm best}$ which determines the YSOs best-fit parameters. A set of `good' fits is then defined as those with $\chi_{\nu}^{2} - \chi^{2}_{\rm best} < 3 \times n_{\rm data}$. 
The range in these model parameters is used to test the sensitivity of the fit to variation in the input parameters and ultimately quantify the error in the best-fit values. Table~\ref{tab:ngc6822_YSOproperties} lists the best-fit value, its median absolute deviation of each parameter of the good fits for each source. For the \citetalias{Robitaille2017} models these parameters were determined from the most probable model set, which provides the most reasonable fits to the data.

As our fitting procedure involves comparing each source to a large set of model templates, it is possible that some of the combinations of parameters may be nonphysical. We filter out any YSO model outside of the parameter space covered by the PARSEC evolutionary tracks \citep{Chen2015,Tang2014}. Furthermore, if several models with vastly divergent properties provide a good fit we may not be able to produce strong constraints on the parameters and hence the nature of the object. In such a situation, we determine which model is most likely based on the complexity required to fit the data and the fraction of models with those properties that provide a good fit to the object's SED.

In some instances, the SEDs of the YSO candidates are adversely affected by the size of the physical aperture at the distance of NGC 6822. Consequently, obtaining a  high-quality SED fit to the data by the theoretical YSO models is challenging.
Incomplete sampling of the SEDs, a bad data point, or unresolved protoclusters masquerading as a single object in environments of illuminated dust may produce poor fits to sources likely to be a YSO.
For many sources the fit is poorly constrained at 8 $\mu$m (even when the PAH correction method is employed) due to strong contamination by the 7.7 $\mu$m PAH emission complex. This is also complicated by the models having a single dust composition and parameters optimised for solar metallicities. For instance, in metal-poor star formation the silicate dust is likely to be oxygen deficient \citep{Jones2012}, which may also affect the fit to the IRAC 8.0 $\mu$m data point.  

\begin{figure*}
    \centering
    \includegraphics[width=0.49\textwidth]{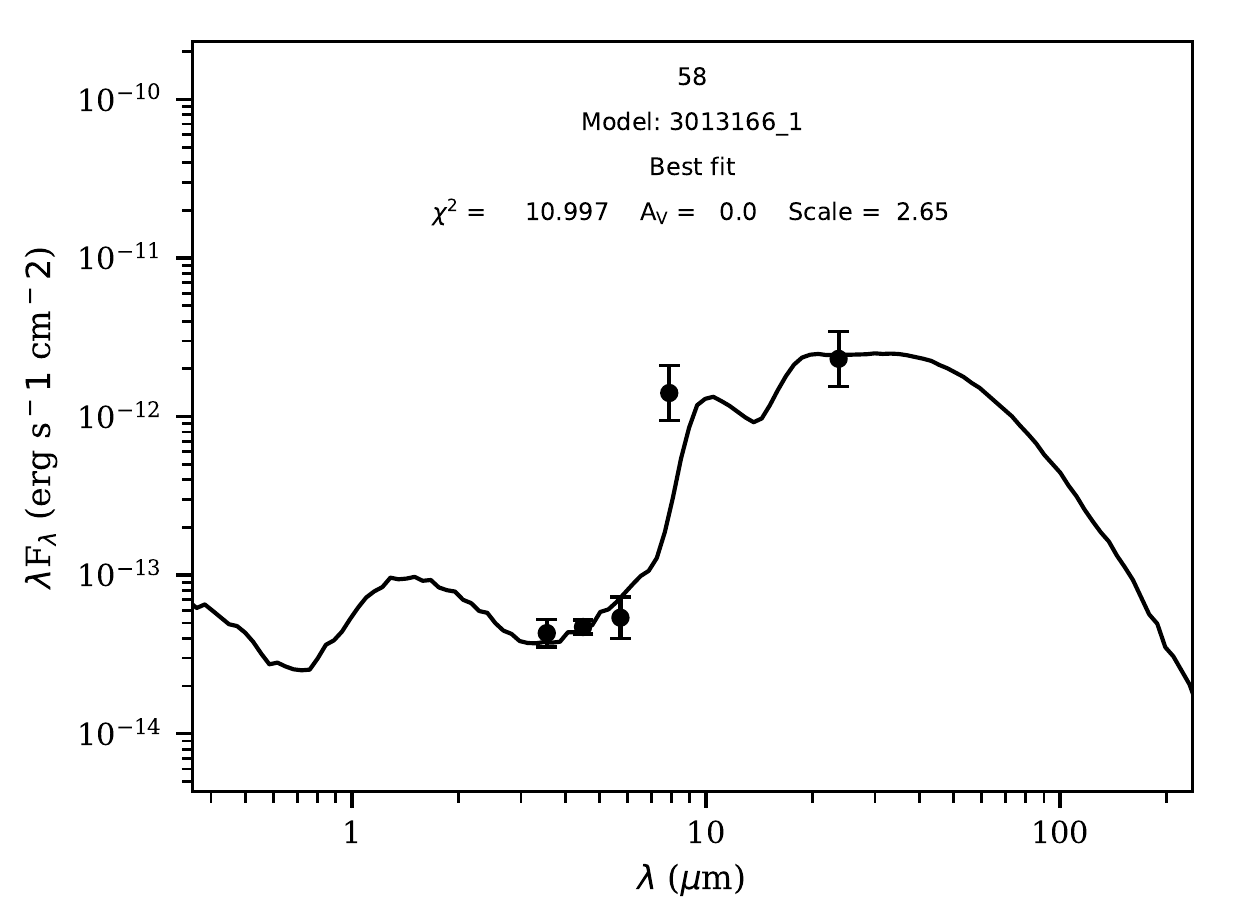}
    \includegraphics[width=0.49\textwidth]{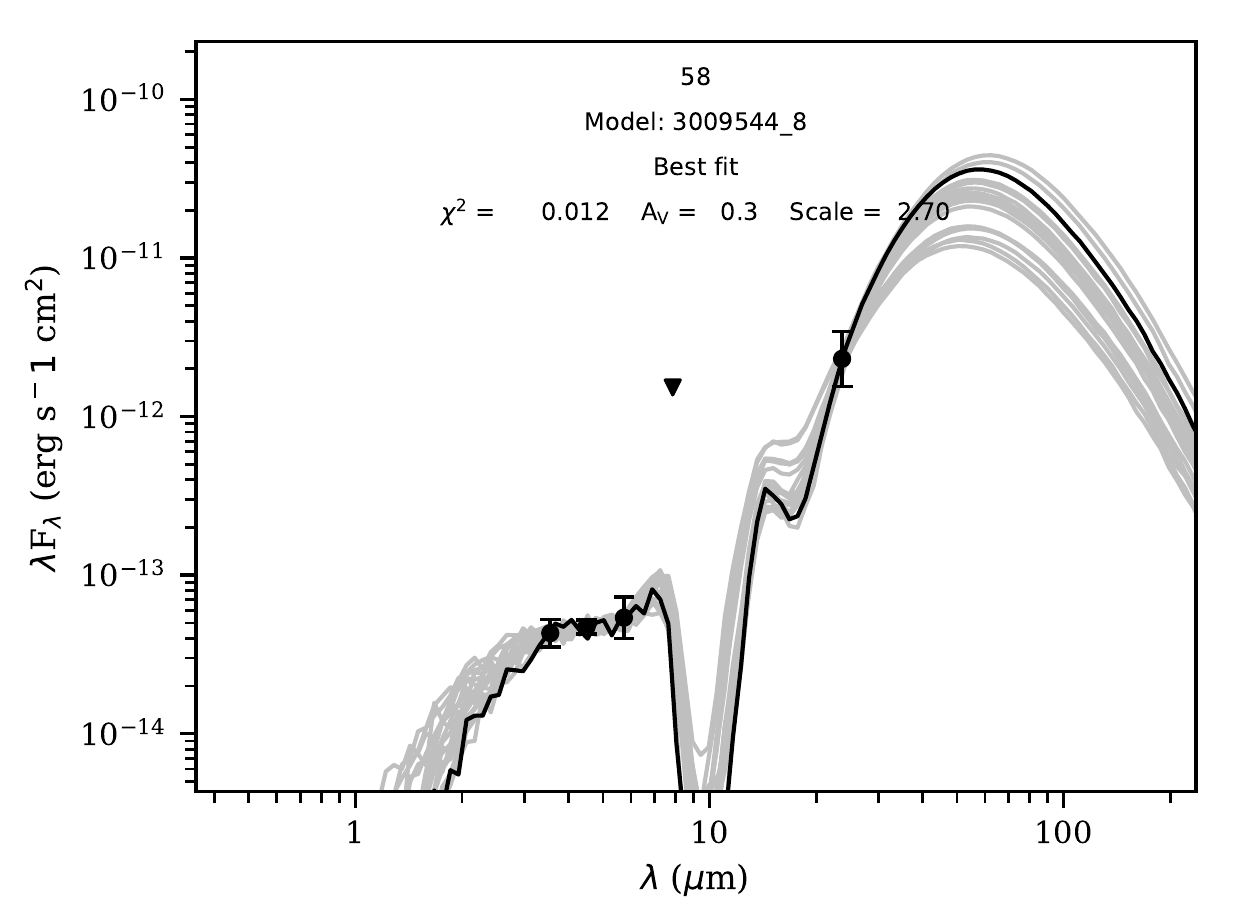}
    \includegraphics[width=0.49\textwidth]{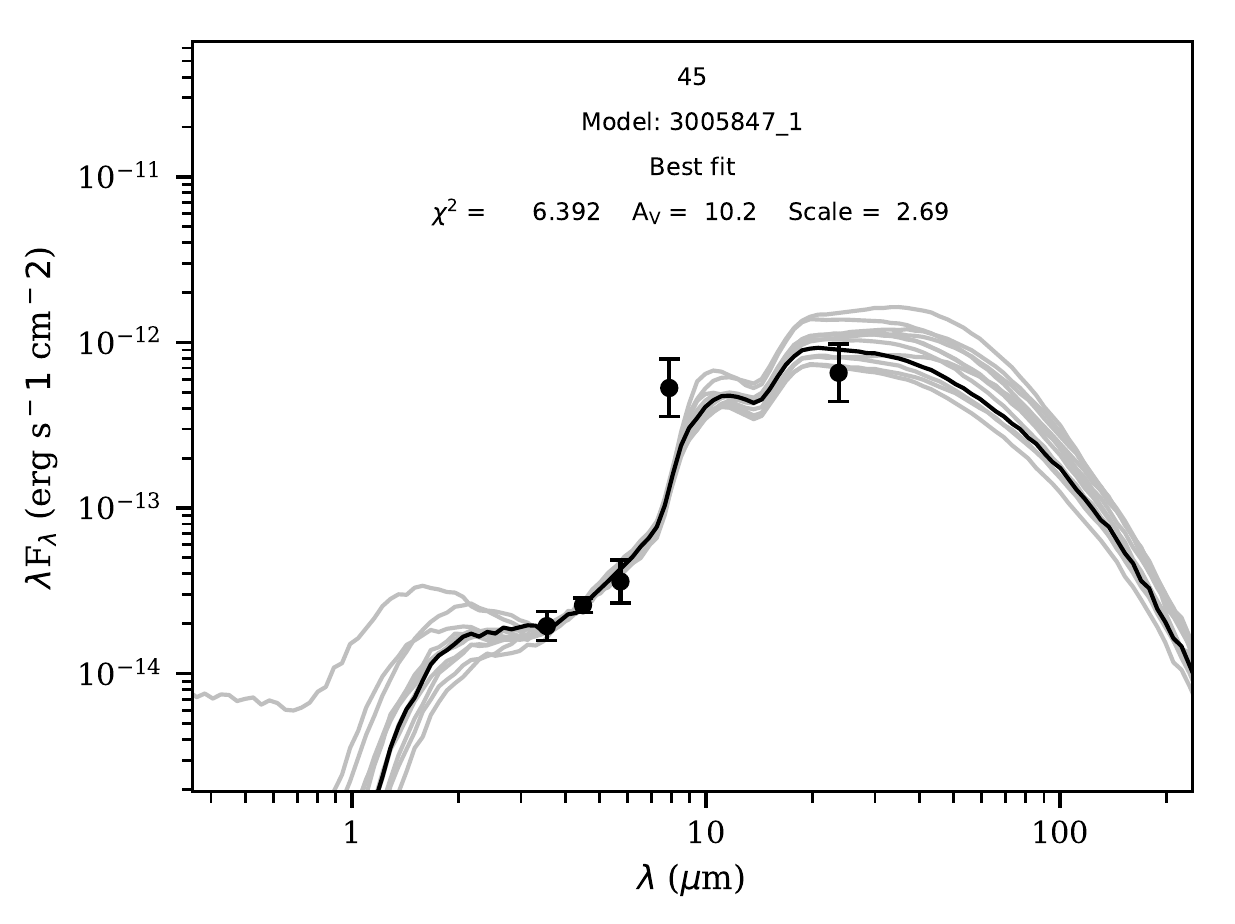}
    \includegraphics[width=0.49\textwidth]{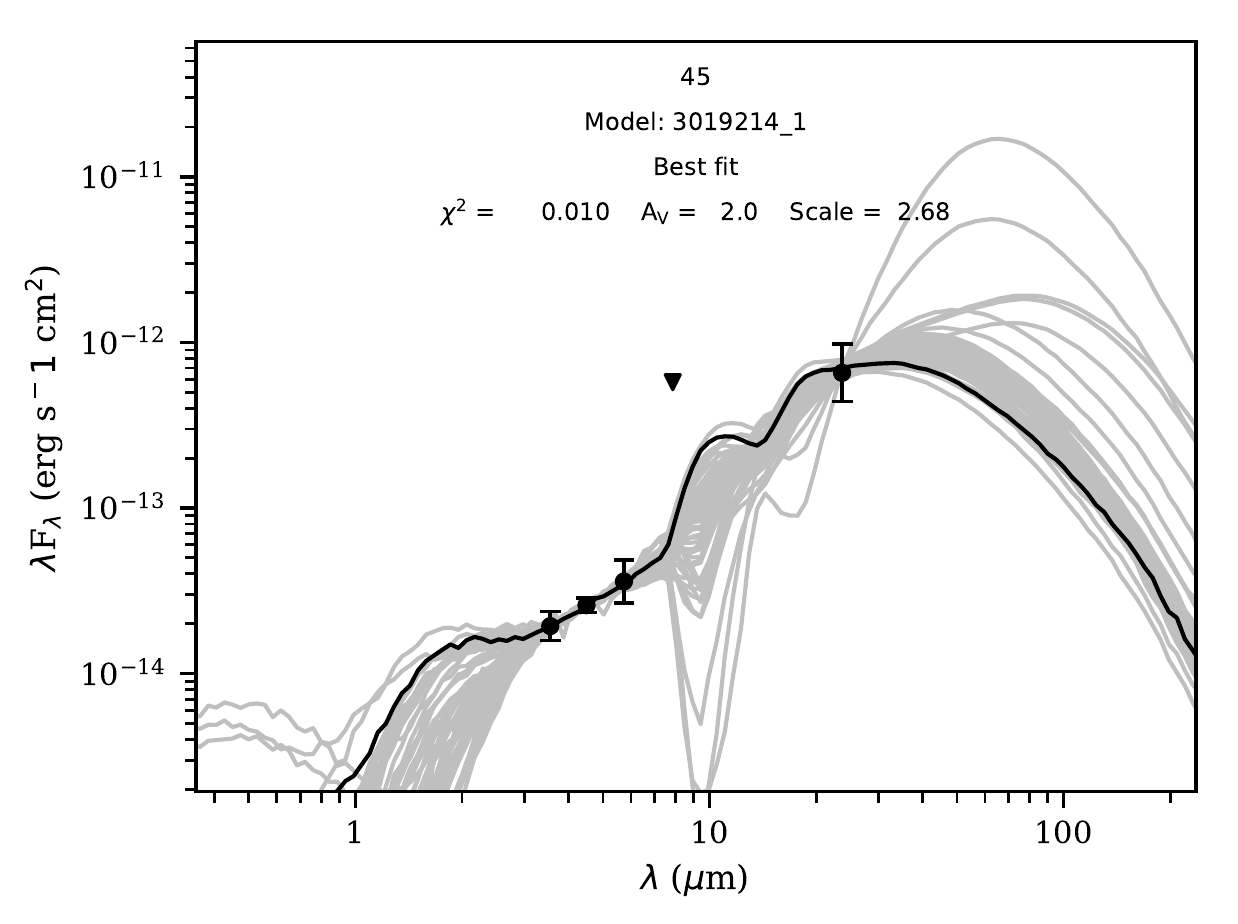}    
    \caption{Comparison of the best-fits with the~\protect\citetalias{Robitaille2006} models to two YSO candidates using the PAH correction method (left) to fits using the upper limit method for the 8 $\mu$m data (right). SEDs of sources with PAHs show a dip at 4.5 $\mu$m; the depth of the feature depends on the strength of the PAH emission. Symbols as in Figure~\ref{fig:SEDmodelfit}. } 
    \label{fig:SEDmodelfit_ULvLargeErr}
\end{figure*}

To quantify how the 8 $\mu$m data affect the quality of the fits for the YSO candidates, we compare the SED fits where the 8 $\mu$m data is treated as an upper limit to fits obtained with increased error bars. Figure~\ref{fig:SEDmodelfit_ULvLargeErr} compares the SED fits using both methods for two example sources. For Source 58 the difference in the best-fit model properties using the two methods is appreciable, whilst little change is seen between the model parameters for Source 45.  In each instance, fitting the data points where the 8 $\mu$m fluxes are considered an upper limit results in a substantial improvement to the quality of the fit and a lower YSO mass estimate.  Consequently, all sources flagged as contaminated by PAHs in the [3.6]--[4.5] vs.~[4.5]--[5.8] CCD have their 8 $\mu$m flux set as an upper limit in the SED fitter input. This method is not appropriate if the flux at 24 $\mu$m is unconstrained. 


\begin{figure*}
    \centering
    \vspace{0 cm}
    \includegraphics[width=0.49\textwidth]{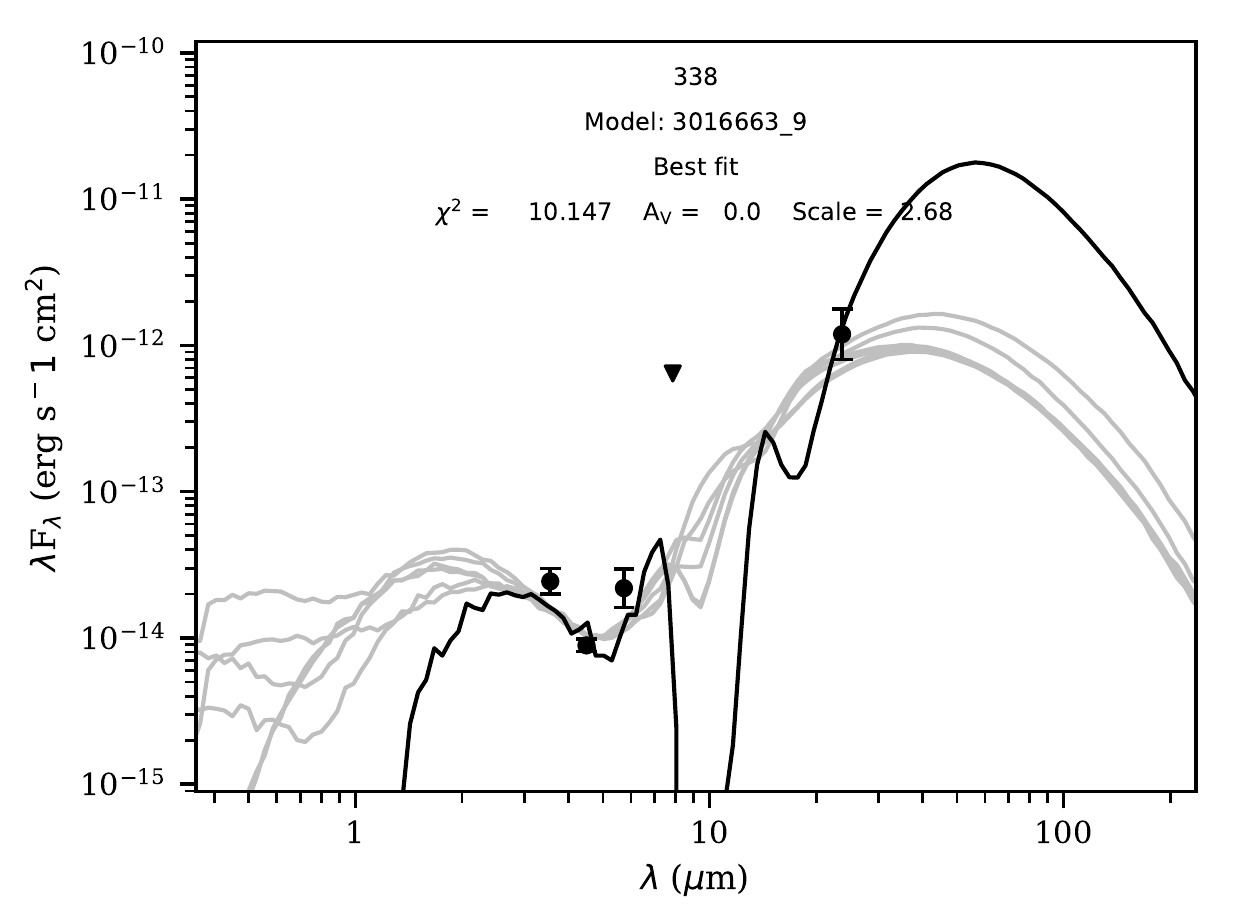}
    \includegraphics[width=0.49\textwidth]{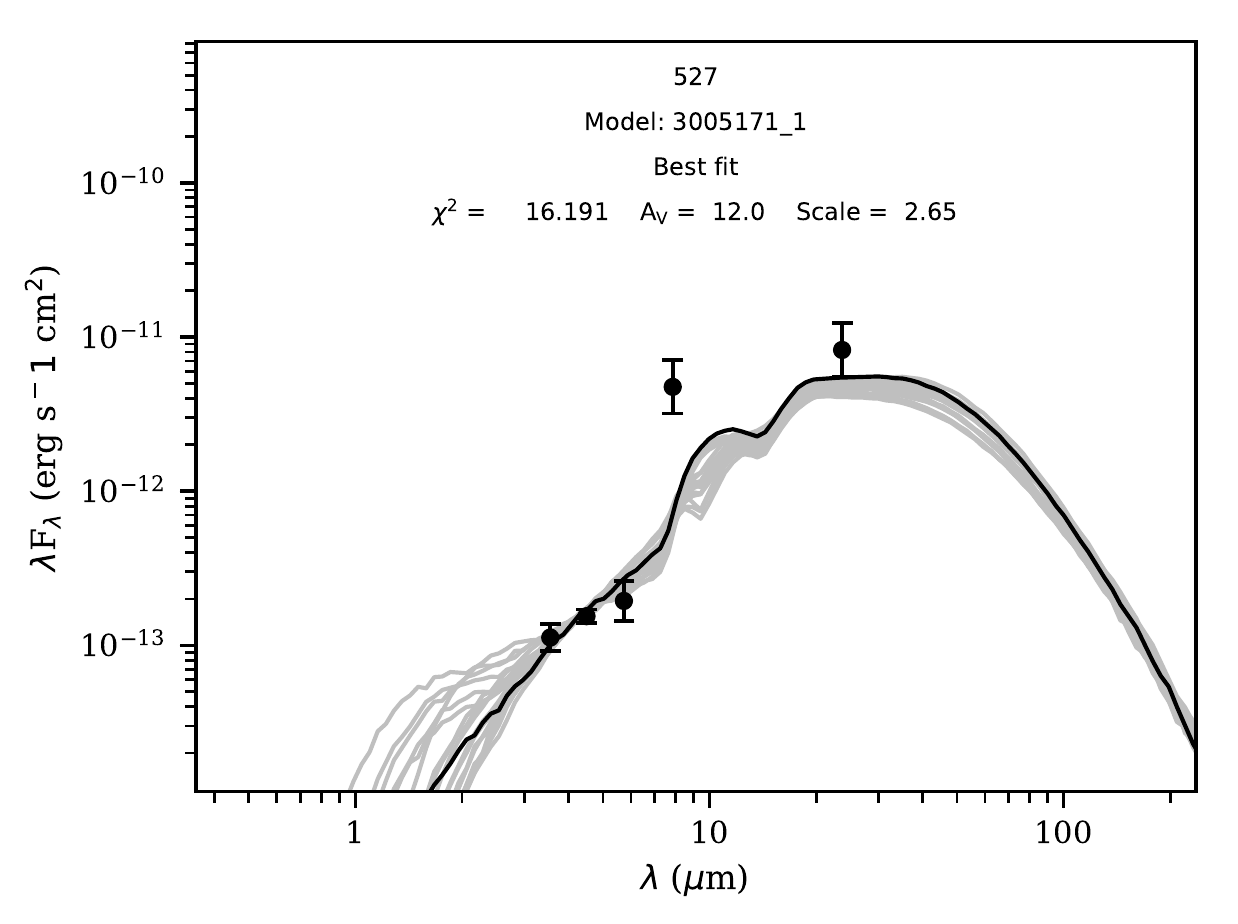}
    \caption{Example SEDs for YSO candidates which have poorer fits to the data by the \citetalias{Robitaille2006} models, thus are considered medium-confidence YSO's (see text). For source 338 severe PAH contamination affects the quality of the fit. Source 527 is classified as a H{\sc ii} region by \citet{HernandezMartinez2009}. Symbols as in Figure~\ref{fig:SEDmodelfit}. } 
    \label{fig:SEDmodelfit_badFit_probYSO}
\end{figure*}

It is probable that some legitimate YSO sources have been excluded from our `high-confidence' list due to their modest YSO model fits. Examples of `medium-confidence' YSOs with poor SED fits are given in Figure~\ref{fig:SEDmodelfit_badFit_probYSO}.   We consider sources to be `medium-confidence YSOs' if they  have a high YSO colour score and an SED shape indicative of a YSO \citep{Lada1987}. These sources typically have a $\chi_{\nu, {\rm best}}^{2}$ in the range 10 to 15. The 13 YSO-candidates which have both good fits by the YSO models and stellar photosphere models are also assigned to this category, leaving 105 high-confidence and 88 medium-confidence sources. 

\section{Results}
\label{sec:results}

\subsection{The physical properties of the high-confidence YSO candidates}
\label{sec:YSOmodelresults}

The results of the \citetalias{Robitaille2006} SED model fits to the high-confidence YSO candidates are given in Table~\ref{tab:ngc6822_YSOproperties}. Only selected physical parameters of the best-fit models are listed to prevent over-interpretation. These are: the total luminosity ($L_{\star }$ and stellar mass (${\rm M}_{*}$), as well as their median absolute deviations. The evolutionary stage, number of well-fit models, and the best fit $\chi_{\nu}^{2}$ is also given.  
For comparison, the best \citetalias{Robitaille2017} model set (which outlines the model components present), the number of best-fit models from the most-likely model set, the best-fit stellar radius ($R_{\star}$),  effective temperature  ($T_{\rm eff}$), and the luminosity for the high-confidence YSO candidates are listed in Table~\ref{tab:ngc6822_YSOproperties_2017}. 

Figure~\ref{fig:SEDmodelfit} shows the difference in the best-fit models to three sources using the  \citetalias{Robitaille2006} and \citetalias{Robitaille2017} model grid. These three sources represent the Stages I--III in a YSOs evolution. Using both sets of models, which have their own advantages and drawbacks, allows us to draw comparisons to previous works using the \citetalias{Robitaille2006} models, place stronger constraints on the error associated with the YSO properties, and restrict correlations between parameters which were present in the \citetalias{Robitaille2006} models. 
For instance, the \citetalias{Robitaille2006} values for the envelope accretion rate and disk mass are used to determine an evolutionary stage, these values can have large uncertainties and may correlate with the stellar mass, whilst the \citetalias{Robitaille2017} models take a Bayesian approach, and the model components present in the best set  determines the evolutionary stage of the source. Source 471 shown in the middle panel of Figure~\ref{fig:SEDmodelfit} is a good example of why one model set was not preferred to the other; it appears to have very different SED fits using the two sets of models, however both YSO evolutionary stages derived from the fitting are in agreement with each other, increasing the confidence in our derived pre-main sequence evolutionary stage.

\begin{figure}
    \centering
    \vspace{0 cm}
    \includegraphics[width=0.49\textwidth]{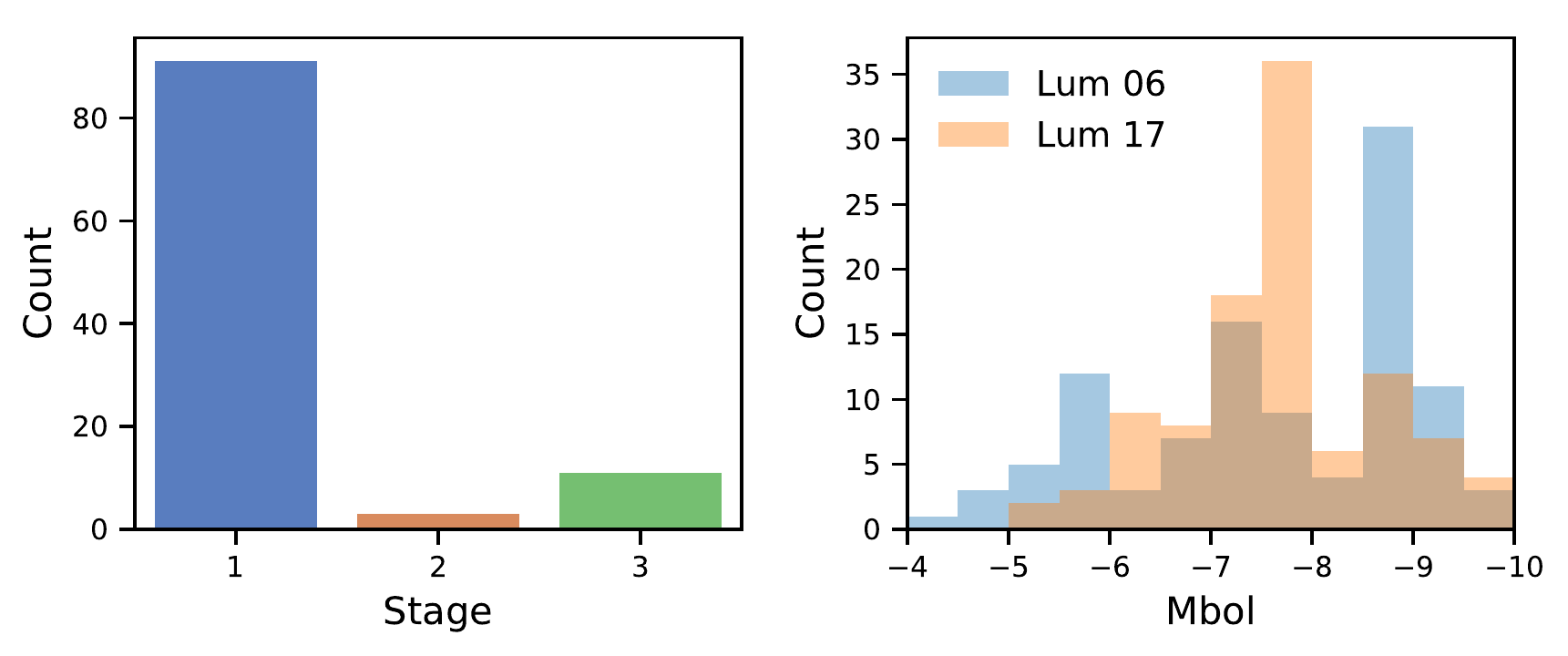}
    \includegraphics[width=0.49\textwidth]{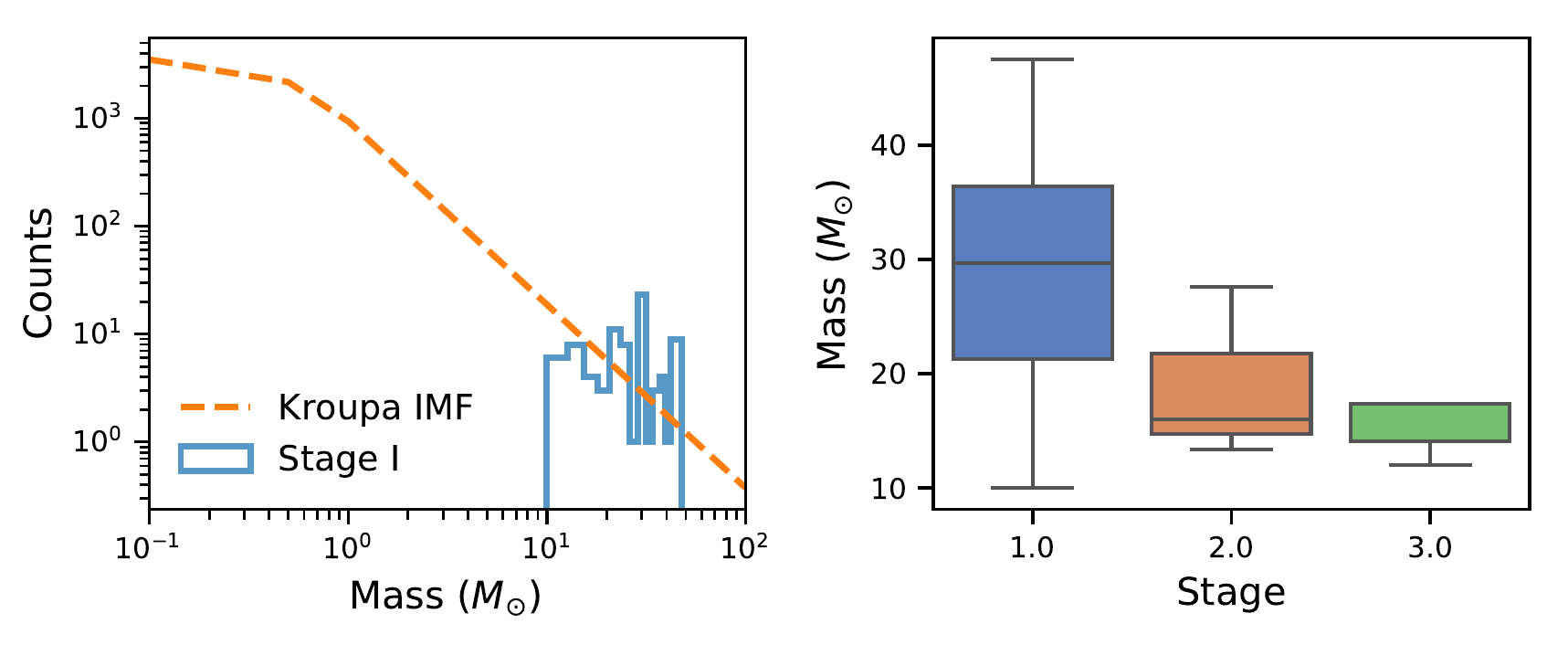}
    \caption{Histogram of the evolutionary stage and bolometric luminosity for the best-fit YSO candidates. The stellar mass distribution of the Stage I YSO candidates fit with a \citet{Kroupa2002} IMF is shown in the lower left. The mass distribution for each stage is shown in the lower right.} 
    \label{fig:HistogramStageMass}
\end{figure}

The YSO fitting provides estimates of the YSOs' evolutionary stage. Embedded Stage I sources have $\dot{M}_{{\rm env}} > 10^{-6} ~M_{\star } {\rm yr}^{-1}$; the more evolved Stage II sources have $\dot{M}_{{\rm env}} < 10^{-6} ~M_{\star }{\rm yr}^{-1}$ and an optically thick disk  $M_{\rm disk}/M_{\star } > 10^{-6}$; Stage III YSOs in the later stages of pre-main-sequence evolution have $M_{\rm disk}/M_{\star } < 10^{-6}$ and $\dot{M}_{{\rm env}} < 10^{-6} M_{\star }{\rm yr}^{-1}$. 
Figure~\ref{fig:HistogramStageMass} shows the distribution of evolutionary stages for our high-reliability YSO candidates.  In total, 91, 3, and 11 stars in NGC 6822 are classified as Stage I, II, and III YSO candidates, respectively.  For the \citetalias{Robitaille2017} models we use the presence of an envelope and/or disk in the best-fitting model set to estimate the YSOs evolutionary stage. 
Our sample preferentially detects young embedded (Stage I) sources, compared to Stage II and III sources as younger sources are bright in the mid-IR compared to YSOs identified via blue tracers (e.g.~UV, H$\alpha$). Generally the stage returned by both model sets is consistent, however for Source 147 shown in Figure~\ref{fig:SEDmodelfit_17v06}, the best-fit \citetalias{Robitaille2017} model-set was a star plus ambient ISM rather than a Stage I object returned using the \citetalias{Robitaille2006} models. 

In NGC 6822, our YSO candidates are probably embedded star-forming clusters, as resolution limitations prevent us isolating individual sources. These clusters will have a range of YSO properties, resulting in the blending of the individual model parameters. For instance, for the circumstellar dust geometry, the disk-flaring angle and scale-height are poorly constrained. This can affect the derived disk mass, envelope accretion rate, and hence the evolutionary stage of the source. Given the limitations of our data, and given that the most-massive and most luminous source in any unresolved cluster dominates the total radiative output, treating the {\em Spitzer} YSO candidates as single massive objects is a reasonable approximation and the aggregate properties listed in Tables~\ref{tab:ngc6822_YSOproperties} and \ref{tab:ngc6822_YSOproperties_2017} are a good estimation.

The stellar luminosity is the most robust parameter obtained from the model fits. From this and the effective temperature ($T_{\rm eff}$), a stellar mass ($M_{\star }$) and age ($t_{\star }$) can be obtained assuming a pre-main-sequence evolutionary track. For the \citetalias{Robitaille2006} models this is pre-computed, whilst the \citetalias{Robitaille2017} models allow the user to specify their own evolutionary track. For these models we must also calculate the YSO's luminosity using the Stefan-Boltzmann law, and the best-fit stellar radius ($R_{\star}$) and effective temperature. The total luminosity of the well-fit YSOs is 1.6 $\pm$ 0.2 $\times 10^{7}$ $L_{\odot}$. 
Figure~\ref{fig:HistogramStageMass} shows the derived luminosity from the \citetalias{Robitaille2006} and  \citetalias{Robitaille2017} models and mass distribution of our high-confidence YSO candidates derived from the SED fitting.  In general, the luminosities derived using the \citetalias{Robitaille2017} models are lower than the \citetalias{Robitaille2006} models. This is due to the better treatment of emission at long-wavelengths in the \citetalias{Robitaille2017} models and the improved sampling in parameter space.
Approximately 25~\% of our sample have an average mass greater than 15$M_{\odot}$. From the shape of the histogram we expect our sample to have the highest completeness for the 15 -- 35 $M_{\odot}$ mass range. Our sample of YSOs is incomplete for low-mass sources due to our selection criteria employed to produce a reliable sample of YSO and limitations in sensitivity and resolution of our data. 
High-mass sources may also be missing, as multiple high-mass stars in proto-clusters will be detected as a single massive source within the IRAC and MIPS PSF. 
We therefore expect our {\em Spitzer} data for each source to encompass multiple YSOs at similar evolutionary stages, with a dominant component arising from the most massive source in the proto-cluster \citep[e.g][]{Oliveira2013, Ward2017}. YSO model fits to these unresolved clusters overestimate the mass of the most massive YSO, but underestimate the total mass of all the YSOs in the beam \citep{Chen2010}. Thus the mass estimates have large uncertainties and should be viewed as a lower limit as we assume single sources. 

\begin{table*}
\centering
 \caption{Physical Parameters of YSO Candidates fit with the~\protect\citepalias{Robitaille2006} models. Only a portion of this table is shown here to demonstrate its form and content. A machine-readable version of the full table is available online.}
 \label{tab:ngc6822_YSOproperties}
\centering
 \begin{tabular}{@{}cccccccccc@{}}
   \hline
   \hline
   Source   &    R.A.  &    Dec.  &  	$\chi^2$  & 	nfits  & 	${\rm M}_{*}$  & 	err. ${\rm M}_{*}$   & 	$L_{\star }$  & 	 err. $L_{\star }$  & 	  Stage \\	
           &           &           &	   &      &   ${\rm M}_{\odot}$ &   ${\rm M}_{\odot}$ &   ${\rm L}_{\odot}$  &   ${\rm L}_{\odot}$ &  	\\	
\hline             
10 & 296.0626 & -14.820 & 7.00 & 744 & 14 & 1 & 17300 & 5400  & 3 \\
16 & 296.0697 & -14.839 & 4.04 & 713 & 10 & 5 & 18600 & 7250  & 1 \\
20 & 296.0714 & -14.888 & 7.03 & 1358 & 12 & 1 & 10900 & 5090  & 3 \\
25 & 296.0775 & -14.881 & 0.70 & 1782 & 16 & 4 & 15400 & 30100 & 1 \\
28 & 296.0806 & -14.964 & 1.52 & 2470 & 20 & 6 & 73100 & 54400 & 1 \\
  \hline
 \end{tabular}
\end{table*}

\begin{table*}
\centering
 \caption{Physical Parameters of YSO Candidates fit with the~\protect\citepalias{Robitaille2017} models. Only a portion of this table is shown here to demonstrate its form and content. A machine-readable version of the full table is available online.}
 \label{tab:ngc6822_YSOproperties_2017}
\centering
 \begin{tabular}{@{}ccccccccccc@{}}
   \hline
   \hline
   Source  &    R.A.  &    Dec.  &  Model	& $n_{\rm fits}$ &  $R_{\star}$      &  err. $R_{\star}$   & $T_{\rm eff}$ & err  $T_{\rm eff}$ &  	$L_{\star }$            & 	 err. $L_{\star }$       \\
           &          &          &	Set     &    set         &  ${\rm R}_{\odot}$ &   ${\rm R}_{\odot}$ &  $K$          &   $K$             &  ${\rm L}_{\odot}$  &   ${\rm L}_{\odot}$ \\
\hline             
10 & 296.0626 & -14.820 & s-u-smi & 240  &  28.33 & 16 & 13220 & 2950  & 21990 & 10820 \\ 
16 & 296.0697 & -14.839 & s-u-smi & 172  &   6.46 & 27 & 29860 & 15840 & 29750 & 29920 \\ 
20 & 296.0714 & -14.888 & s-u-smi & 113  &  72.62 & 28 & 6700  & 7320  & 39570 & 12460 \\ 
25 & 296.0775 & -14.881 & s-u-smi & 254  &  29.91 & 20 & 17920 & 3580  & 82770 & 22160 \\ 
28 & 296.0806 & -14.964 & s---smi & 122  &    95 &  31 & 11730 & 1460  & 153880 & 50475 \\
  \hline
 \end{tabular}
\end{table*}

\section{Discussion}
\label{sec:discussion}

\subsection{Spatial Distribution and Comparison to Gas and Dust Tracers}
\label{sec:spatialDist}

\begin{figure*}
    \centering
    \vspace{0 cm}
    \includegraphics[trim=1 40 45 1,clip,width=0.49\textwidth]{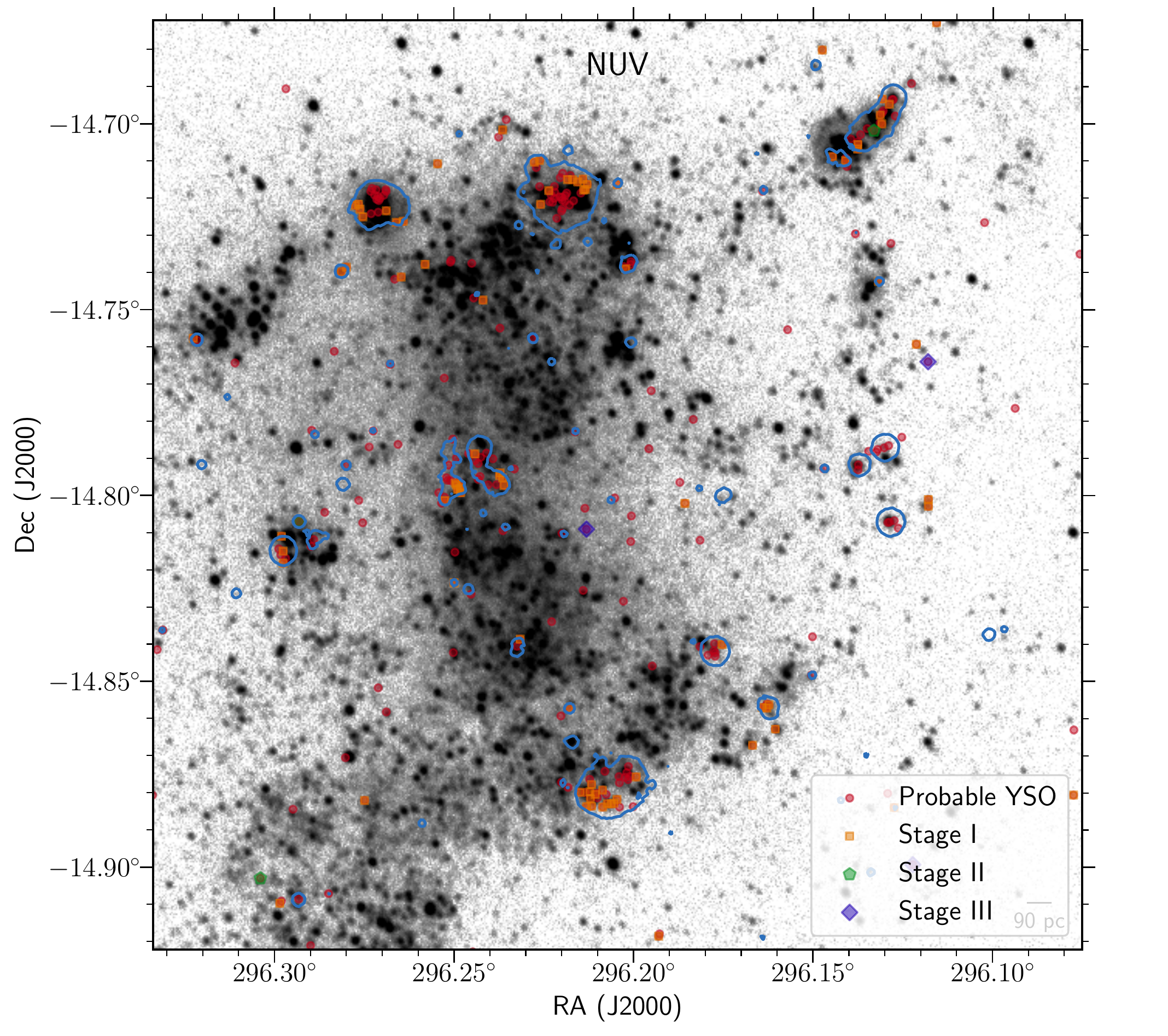}
    \includegraphics[trim=70 40 1 1,clip,width=0.47\textwidth]{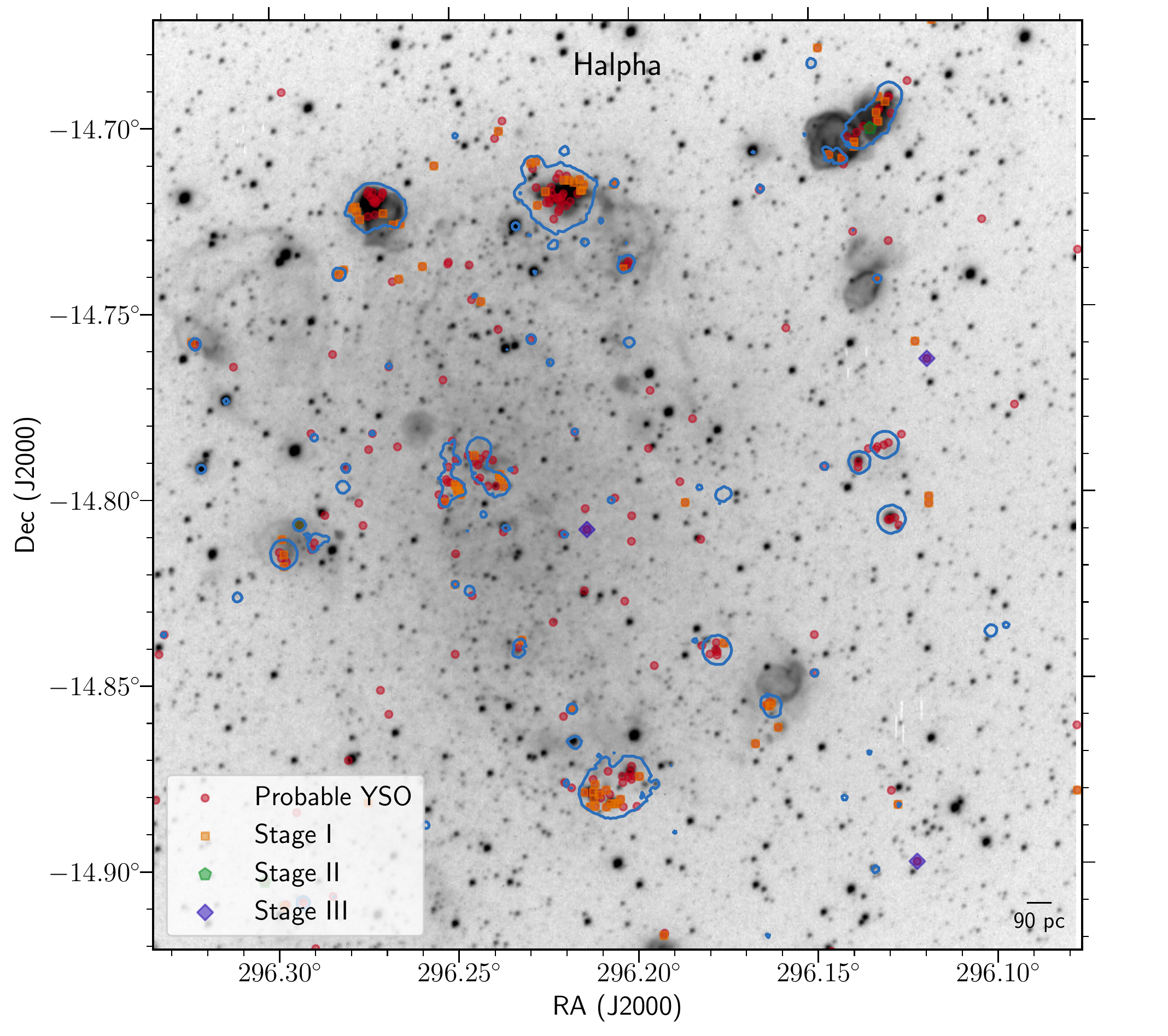}
    \includegraphics[trim=1 1 45 1,clip,width=0.49\textwidth]{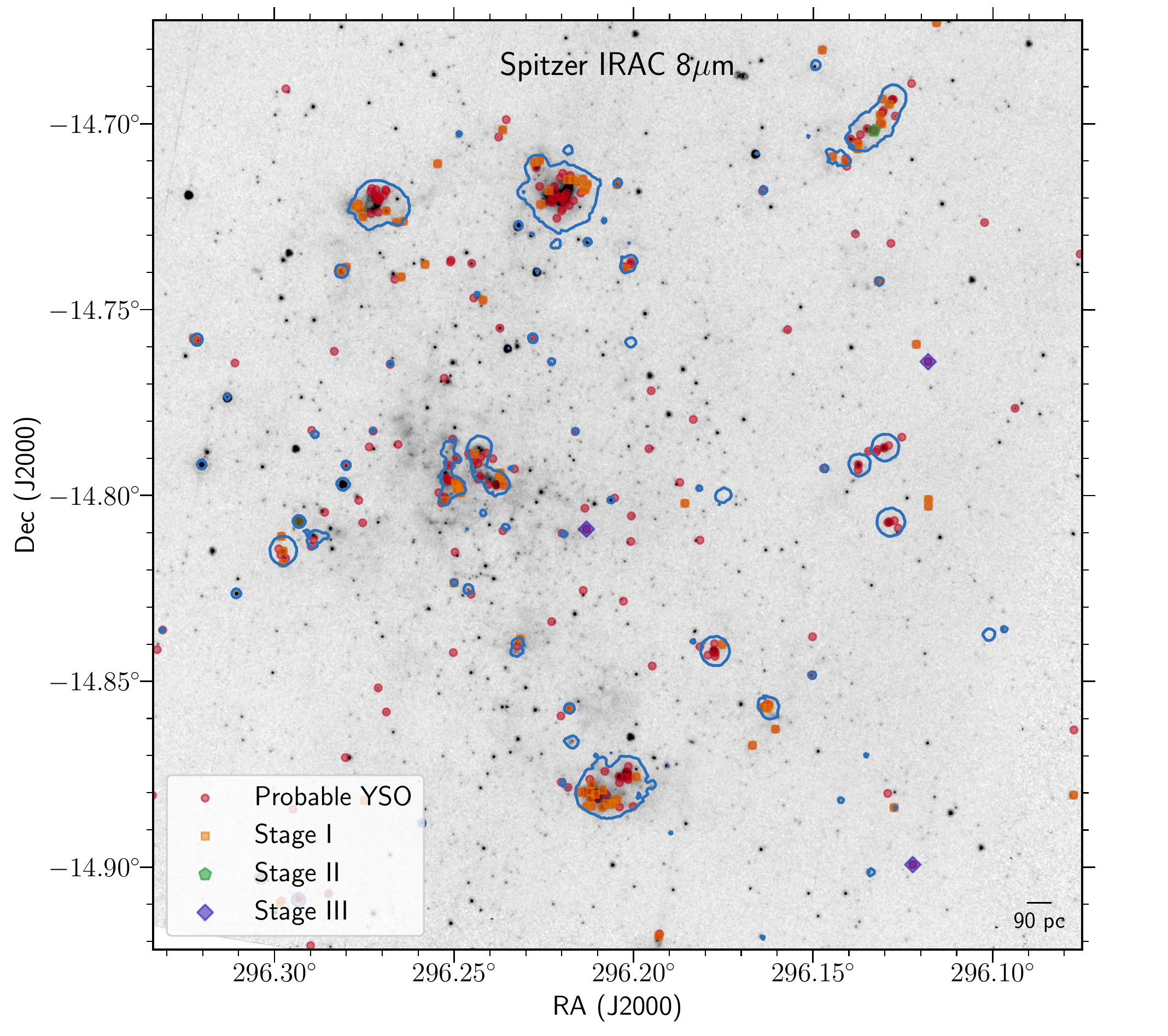}
    \includegraphics[trim=70 1 1 1,clip,width=0.47\textwidth]{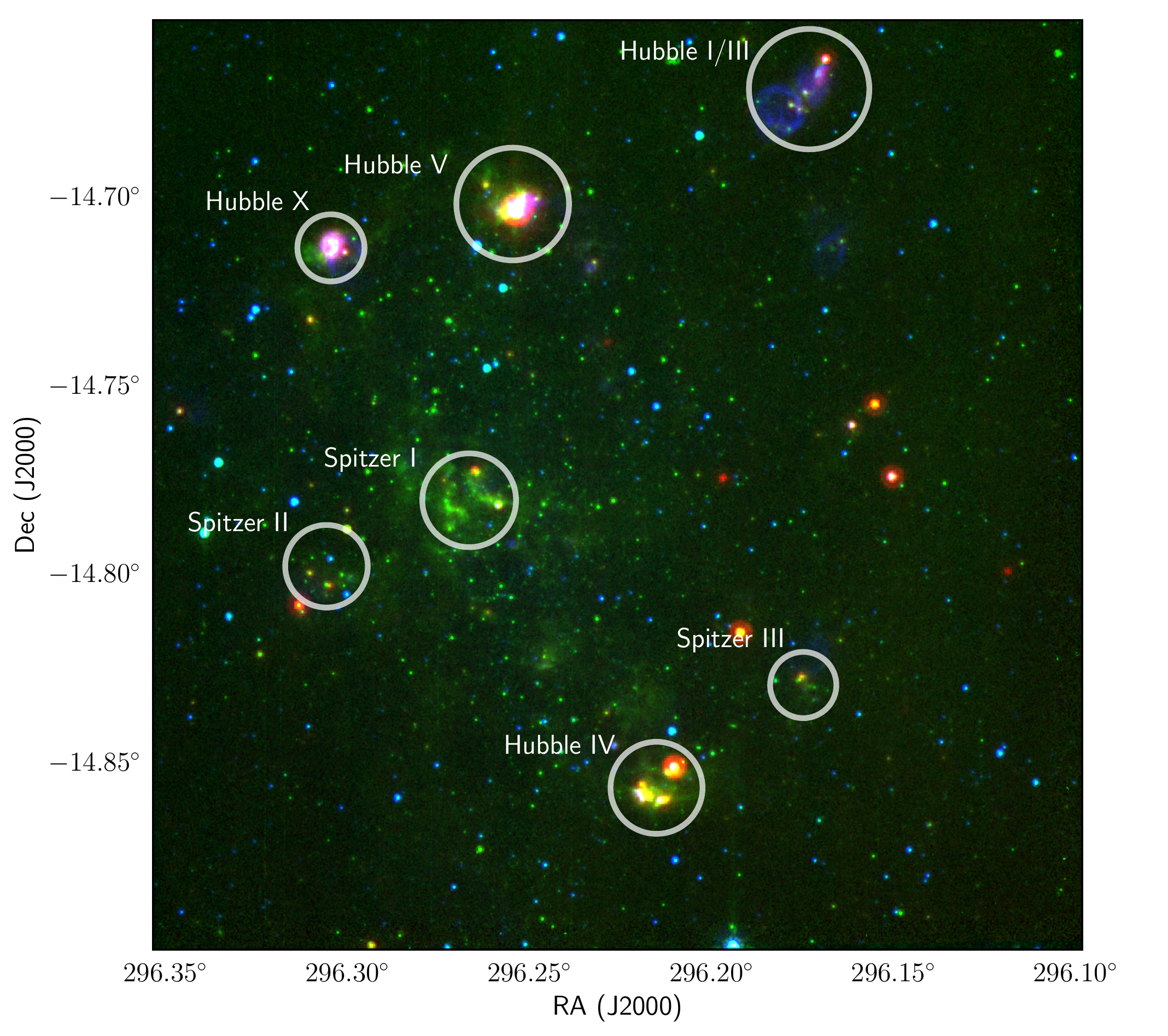}
    \caption{Distribution of the YSO candidates overlayed on grey-scale images of NGC 6822 at  near-UV, H$\alpha$ and 8$\mu$m. The blue contours correspond to a 5$\sigma$ MIPS 24 $\mu$m detection; orange squares show Stage I YSO candidates; green pentagons Stage II candidates and purple diamonds Stage III sources. The `medium-confidence-YSO' candidates are shown as red circles. For reference a three-colour composite image of NGC 6822 combining the MIPS 24$\mu$m (red), IRAC 8$\mu$m (green), and H$\alpha$ (blue) images is shown to the same scale and orientation. } 
    \label{fig:ysoMaps}
\end{figure*}

The spatial distribution of the YSO candidates is overlaid on the near-ultraviolet (near-UV) \citep{Hunter2010, Zhang2012}, H$\alpha$ \citep{Hunter2004} and IRAC 8 $\mu$m \citep{Kennicutt2003} images in Figure~\ref{fig:ysoMaps}. The blue contours in the figure correspond to a 5$\sigma$ MIPS 24 $\mu$m detection. The 8~$\mu$m emission primarily originates from PAH molecules which trace diffuse extended structure in photodissociation regions (PDRs). The 24~$\mu$m surface brightness traces warm dust associated embedded star-formation, whilst the H$\alpha$ and photospheric UV emission reveals unobscured young, massive stars and hence a more advanced epoch of star-formation. NGC 6822 has not been mapped in CO, however there is a close correlation between CO and 8 $\mu$m emission on large scales \citep{Gratier2010, Schruba2017}. Thus the IRAC 8 $\mu$m image is an effective tracer of molecular gas in the galaxy, particularly at low metallicity \citep{Regan2006, Sandstrom2012, Cortzen2019}. 
This emission is relatively high in diffuse regions and low in high-UV environments. 
In NGC 6822 the 24 $\mu$m emission is not co-spatial with the H$\alpha$ emission. 
The YSO candidates tend to be located in the densest regions of 24$\mu$m flux, whilst the H$\alpha$ images show some evidence for YSO candidates located along ridges or in the shells of bubbles. Indicating that star formation in these regions may have been triggered by previous generation of massive star formation. 

There is a strong degree of clustering which coincides with the major star-forming regions: Hubble  X, V, IV, III and I, and two newly discovered embedded star formation regions we dub Spitzer I and II, which are located in the centre and to the centre-east of the bar.  
To quantify the number of star-formation regions in the galaxy, the number of sources associated with each region, and the radius of the semi-major axis for the region, we use the density-based spatial clustering of applications with noise (DBSCAN) clustering algorithm \citep{Ester1996}. This algorithm is very effective for membership determination of arbitrarily-shaped clusters and rejecting outliers without any prior assumptions about the number of clusters or about the stellar distribution. Once we have estimated the number of clusters we then calculate their physical parameters. 
Using the 584 colour-selected YSO candidates, we identify seven major star-forming regions in NGC 6822 with the DBSCAN algorithm. Of these, 294 sources were associated with a cluster and the remaining 290 sources are considered a `noise' point. Each cluster must have at least seven sources within the maximum search radius to be considered. 
For comparison, we also determine the number of star-forming clusters in NGC 6822 using the high-confidence and medium-confidence YSO candidates listed in Tables~\ref{tab:ngc6822_HP_YSOtable} and~\ref{tab:ngc6822_YSOtable_prob}. 
Table~\ref{tab:ngc6822_staformingRegions} lists the star-forming cluster properties; the cluster centres are determined from the mean position of all YSOs associated with that cluster and the semi-major axis radius from the outermost high-confidence YSO member from the cluster centre. Over 60 percent of the high-confidence YSO candidates are associated with a cluster, with the remainder mostly distributed along dusty ridges in the bar of NGC 6822.   Massive YSOs in the Magellanic clouds show a similar correlation \citep{Whitney2008, Ochsendorf2016}. 

Seven major star-formation regions were identified in NGC 6822. 
These include three bright regions in H$\alpha$: Hubble IV, Hubble V, and Hubble X, located in the south, north-east and north-west of the bar, respectively. These are young
\Hii~regions powered by massive O and B stars which contain 53, 53, and 33 colour-selected YSO candidates, respectively.
The ring nebula, Hubble III, and the neighbouring Hubble I nebula in the north-west region off from the main bar are identified as one elongated region by the DBSCAN routine. Here, Figure~\ref{fig:ysoMaps} shows that the majority of the {\em Spitzer}-identified YSOs are associated with Hubble I rather than Hubble III. The number of Stage I and II YSOs and the ratio of 24 $\mu$m to H$\alpha$ emission suggests that Hubble I is more active and younger than Hubble III. 
Hubble IV and V are thought to be even younger than Hubble I, III, and X \citep{Schruba2017}, with more compact CO morphologies and higher ratios of embedded to exposed SFR tracers. Our star counts of YSO candidates confirm this. 

Three new regions of embedded star formation were also discovered:
Spitzer I is a new area of active star formation. It has substantial 8~$\mu$m emission, some 24 $\mu$m emission, but is faint in H$\alpha$ and the UV.  This region contains 90 YSO candidates, the highest number in NGC 6822. Together with Hubble V, Spitzer I is rich in molecular gas inferred from the 8~\micron\ emission \citep{Sandstrom2012} and is probably the youngest SFR in the galaxy. The high IR flux compared to UV or H$\alpha$ indicates that star formation is on
rise in this region, and it has has yet to reach its peak star-formation activity, similar to the N79 object H72.97-69.39 \citep{Ochsendorf2017}.

Spitzer II has a substantial fraction of YSO candidates, however the number of high-confidence sources is lower than the Hubble regions and Spitzer I. Its properties are comparable to Hubble V and is probably of a similar age. 

Spitzer III is a small region of embedded star-formation containing 11 YSO candidates; the lowest of all the regions identified using DBSCAN. It has a similar H$\alpha$-to-IR flux ratio as Hubble IV, and encompasses the \Hii\ regions KD7--9 \citep{Killen1982}.

Unfortunately, we have no direct measurements of the molecular gas content of the southern part of the galaxy which encompasses Spitzer I, II and III, only the major Hubble regions and the northern section and of NGC 6822 have been observed in the CO(2-1) line \citep{Gratier2010, Schruba2017}.


For the `possible' YSO candidates listed in Table~\ref{tab:ngc6822_YSOtable_poss} identified using the $JHK$, 3.6 $\mu$m and 4.5 $\mu$m fluxes which are not well-fit by a stellar photosphere model, we separate clusters of high density from the rest of the populations via the DBSCAN algorithm. The majority of the sources are associated with the seven star-forming region listed in Table~\ref{tab:ngc6822_staformingRegions}; based on their locations and their SED shapes, these are likely true YSOs at a more advanced stage of formation.  From this population of sources we note two clusters: one at 296.134 -14.789 (identified as a \Hii\ region by \citealt{Hodge1988}) and one located at 296.178 -14.842 with radii of 118 and 31 pc respectively.

\begin{table}
\centering
\caption{Major Star-Formation Regions in NGC 6822.}
\label{tab:ngc6822_staformingRegions}
\scalebox{0.85}{
\begin{tabular}{@{}l@{\ \ \ }ccccc@{}}
   \hline
   \hline
Name	&	Region	&	RA	&	Dec	&	Radius	&	Number of colour  \\    
        &    ID     &       &       &     (pc)  &   selected YSOs  \\   
\hline
Hubble I/III    	&	0	&	296.1338	&	-14.7006	&	140	&	24	\\   
Spitzer III	&	1	&	296.1626	&	-14.8585	&	77	&	11	\\   
Hubble IV	        &	2	&	296.2076	&	-14.8793	&	107	&	53	\\ 
Hubble V        	&	3	&	296.2204	&	-14.7180	&	131	&	53	\\ 
Spitzer I	    &	4	&	296.2459	&	-14.7947	&	109	&	90	\\   
Hubble X        	&	5	&	296.2723	&	-14.7216	&	78	&	33	\\  
Spitzer II	&	6	&	296.2881	&	-14.8059	&	96	&	30	\\   
\hline
\end{tabular}}
\end{table}

\subsection{The current star-formation rate}

To estimate the global star formation rate (SFR) for NGC 6822, we use the YSO counts and their derived mass at each evolutionary stage.  
Figure~\ref{fig:HistogramStageMass} shows a histogram of the measured YSO masses for our sample. The total mass of our 105 high-confidence YSOs is 2810~$M_{\odot}$. This is dominated by the 91 Stage I sources with a mass of 2580~$M_{\odot}$. Assuming our sample is approximately complete at the peak of the distribution, we can fit a \citet{Kroupa2002} initial mass function (IMF) to this peak, in the mass range 20 -- 35~$M_{\odot}$. This IMF is a broken power law where $\alpha = -1.3$ for 0.08 $< M_{\rm star} <$ 0.5 M$_{\odot}$ and $\alpha = -2.3$ for $ M_{\rm star} > 0.5 M_{\odot}$. 
To obtain a total YSO mass in NGC 6822, we then integrate under the IMF for the mass range 0.08 to 50 M$_{\odot}$, resulting in a lower mass limit of 7800~$M_{\odot}$ for our high-confidence Stage I sources. 
The current SFR can be estimated by dividing the total mass of YSO candidates by the formation timescale, assuming this is constant over time. Here we consider only Stage I sources as they trace the most recent star-formation, are of approximately equal age, and these young embedded massive YSOs are preferentially detected by our {\em Spitzer} sample, thus providing the best completeness. 
Following the examples of \citet{Whitney2008, Lada2010, Sewilo2013, Sewilo2019, Carlson2012} we assume a formation timescale of 0.2 Myr  for Stage I YSOs \citep{Lada1999}. This results in a global star-formation rate for NGC 6822 of 0.039 $\pm$ 0.012 $M_{\odot} yr^{-1}$. 

Our global SFR estimate is in good agreement (considering the uncertainties) with current SFR estimates of 0.04 $M_{\odot} yr^{-1}$ derived from the stellar population of NGC 6822 \citep{Gallart1996} and 0.02 $M_{\odot} yr^{-1}$ based on integrated 24 $\mu$m emission by \citet{Efremova2011}, but higher than the UV SF estimates of 0.014  $M_{\odot} yr^{-1}$, which trace SF over the last 100 Myr. SFR estimates based on H$\alpha$ emission which traces the hottest, most massive stars range between 0.01 -- 0.016 $M_{\odot} yr^{-1}$ \citep{Hunter2004, Cannon2006} are also lower than our estimated SFR. In general, SFR measured from H$\alpha$ and 24 $\mu$m integrated light tend to be lower than SFR estimates from YSO counts \citep{Chen2010}. The SFRs calculated from star counts of Stage I objects and from 24 $\mu$m emission depicts an early ($<1$My), embedded phase of star formation; thus the higher value compared to UV and H$\alpha$ tracers may suggest a recent increase in the SFR in NGC 6822. 
This is surprising as major mergers or other dynamical interactions are usually required to stimulate such an increase in the global SFR of a galaxy on a timescales of $\sim$1 Myr. 

NGC 6822 is thought to have a high SFR-to-CO ratio typical of dwarf galaxies \citep{Lee2009}, however the four prominent star-forming complexes in NGC 6822 observed by ALMA may have a relatively short depletion time compared to the rest of the galaxy \citep{Schruba2017}.  This could be explained if NGC 6822 has experienced a recent starburst, with feedback dispersing the gas reservoirs.   \citet{Schruba2017} dismiss this scenario as unfeasible, as there is no evidence that NGC 6822 has recently undergone a burst of star formation - the global SFR of NGC 6822 is thought to have be constant over the last 400 Myr \citep{Efremova2011}. We also see no evidence for a starburst, however our results indicate lots of current activity and we obtain a higher rate of star formation, by a factor of two, compared to UV and H$\alpha$ measurements. 


The Magellanic Clouds are a pair of metal-poor interacting galaxies. Compared to the SMC which has a SFR of $\sim$0.06 $M_{\odot} yr^{-1}$ \citep{Wilke2004, Sewilo2013} and a high SFR-to-CO ratio typical of dwarf galaxies, the SFR of NGC 6822 is comparable to that of the SMC if we account for the difference in mass between the galaxies. The SMC and NGC 6822 are currently more active than the LMC, which has a SFR of 0.06 $M_{\odot} yr^{-1}$ \citep{Whitney2008} and a total mass 5-10 times greater \citep{Monachesi2012}.

\section{Conclusions}
\label{sec:conclusion}

We present a comprehensive study of the massive young stellar population in NGC 6822 as observed with {\em Spitzer}. Metal-poor YSO candidates were identified via CMD examination and SED fitting. Using six mid-IR colour-cuts we find over 500 YSO candidates in seven massive star-formation regions. This is the first catalogue of young embedded stars in the process of formation identified in the galaxy. Removing known contaminants and through multi-wavelength SED fitting to the data with the \citetalias{Robitaille2006} an \citetalias{Robitaille2017} models, we compile a robust inventory of 105 high-confidence and 88 medium-confidence YSO candidates. The majority of these sources are Stage I YSOs with an accreting envelope in the initial stages of formation.  For these sources we also determine the YSO mass and bolometric luminosity from the effective temperature and radius.  Fitting a \citet{Kroupa2002} IMF to the mass distribution of the Stage I YSOs, which are actively forming, we determine a global star-formation rate for NGC 6822 of 0.039 $\pm$ 0.012 $M_{\odot} yr^{-1}$. This is likely a lower limit with a high uncertainty as we expect most of our YSO candidates to be resolved into clusters of young stars with higher-resolution mid-IR observations.

The {\em Spitzer} catalogue of YSOs in NGC 6822 presented here is incomplete. Our data is only sensitive to massive sources at early evolutionary stages. The addition of higher-sensitivity data with improved resolution would resolve protostar clusters and detect stars forming at lower masses, revealing new star formation sites. 
Future studies with the James Webb Space Telescope (JWST) would unveil more YSO candidates at all stages of evolution and provide better sampling of the SED for each source, improving their identification and evolutionary classification \citep{Jones2017a}.
Improved spatial resolution would also allow better comparisons to gas tracers and to examine substuctures and filaments at smaller spatial scales in these massive star-forming complexes. 

We compare the distribution of the YSO candidates with respect to the large-scale gas and dust emission. The YSO sources have a clumpy distribution, with the majority clustering into seven active high-mass star-formation regions which are strongly correlated with the 8 and 24 $\mu$m emission from PAHs and warm dust.  The clustered distributions also indicate contamination from AGB stars and background galaxies is low. 
The majority of star-formation in NGC 6822 was thought to take place in four prominent H\,{\sc ii} regions: Hubble I/III, IV, V and X.  We identify a new high-mass star formation region, Spitzer I, which hosts the highest number of embedded YSO candidates in NGC 6822. The properties of Spitzer I suggests it is younger and more active than the other prominent well-studied star-formation regions in the galaxy. 

\section*{Acknowledgements}
OCJ and MR has received funding from the EUs Horizon 2020 programme under the Marie Sklodowska-Curie grant agreement No 665593 awarded to the STFC. ASH and MM acknowledge support from NASA grant NNX14AN06G. This research made use of Astropy,\footnote{http://www.astropy.org} a community-developed core Python package for Astronomy \citep{Astropy2013}; APLpy, an open-source plotting package for Python \citep{aplpy2012}; and the SIMBAD database, operated at CDS, Strasbourg, France \citep{SIMBAD}.




\input{journaldefs}

\bibliographystyle{mnras}
\bibliography{libby} 








\bsp	
\label{lastpage}
\end{document}

%% file: journaldefs.tex

\def\aj{AJ}					
\def\actaa{Acta Astron.}                        
\def\araa{ARA\&A}				
\def\apj{ApJ}					
\def\apjl{ApJL}					
\def\apjs{ApJS}					
\def\ao{Appl.~Opt.}				
\def\apss{Ap\&SS}				
\def\aap{A\&A}					
\def\aapr{A\&A~Rev.}				
\def\aaps{A\&AS}				
\def\azh{AZh}					
\def\baas{BAAS}					
\def\jrasc{JRASC}				
\def\memras{MmRAS}				
\def\mnras{MNRAS}				
\def\pra{Phys.~Rev.~A}				
\def\prb{Phys.~Rev.~B}				
\def\prc{Phys.~Rev.~C}				
\def\prd{Phys.~Rev.~D}				
\def\pre{Phys.~Rev.~E}				
\def\prl{Phys.~Rev.~Lett.}			
\def\pasp{PASP}					
\def\pasj{PASJ}					
\def\qjras{QJRAS}				
\def\skytel{S\&T}				
\def\solphys{Sol.~Phys.}			
\def\sovast{Soviet~Ast.}			
\def\ssr{Space~Sci.~Rev.}			
\def\zap{ZAp}					
\def\nat{Nature}				
\def\iaucirc{IAU~Circ.}				
\def\aplett{Astrophys.~Lett.}			
\def\apspr{Astrophys.~Space~Phys.~Res.}		
\def\bain{Bull.~Astron.~Inst.~Netherlands}	
\def\fcp{Fund.~Cosmic~Phys.}			
\def\gca{Geochim.~Cosmochim.~Acta}		
\def\grl{Geophys.~Res.~Lett.}			
\def\jcp{J.~Chem.~Phys.}			
\def\jgr{J.~Geophys.~Res.}			
\def\jqsrt{J.~Quant.~Spec.~Radiat.~Transf.}	
\def\memsai{Mem.~Soc.~Astron.~Italiana}		
\def\nphysa{Nucl.~Phys.~A}			
\def\physrep{Phys.~Rep.}			
\def\physscr{Phys.~Scr}				
\def\planss{Planet.~Space~Sci.}			
\def\procspie{Proc.~SPIE}			
\let\astap=\aap
\let\apjlett=\apjl
\let\apjsupp=\apjs
\let\applopt=\ao
